\documentstyle[aps,multicol,epsf]{revtex}

\def\lot{\leq}
\def\got{\geq}
\def\BSigma{{\bf \Sigma}}
\def\wnc{{ln[c(\x,t)]}}
\def\u{{\bf u}}
\def\k{{\bf k}}
\def\q{{\bf q}}

\def\D{{\bf D}}
\def\v{{\bf v}}

\def\y{{\bf y}}
\def\e{{\bf e}}
\def\g{{\bf g}}
\def\f{{\bf f}}

\def\H{{\bf H}}

\def\cH{{\cal H}}
\def\cL{{\cal L}}
\def\cZ{{\cal Z}}
\def\eps{\epsilon}
\def\teps{{\tilde \epsilon}}

\def\x{{\bf x}}
\def\p{{\bf p}}

\def\r{{\bf r}}
\def\be{\begin{equation}}
\def\ee{\end{equation}}

\def\c{\bf c}

\def\cZ{{\cal  Z}}
\def\rp{\r_\perp}

\def\tu{{\tilde {U}}}

\begin{document}
\title{Non-Hermitian Localization and Population Biology} 
\author{David  R. Nelson and Nadav M.  Shnerb}
\address{ Lyman Laboratory  of Physics, Harvard  University,
Cambridge, MA 02138}
\date{\today}
\maketitle

\begin{abstract}
The time evolution of spatial  fluctuations in  
inhomogeneous $d$-dimensional  
 biological systems is analyzed. A single species  continuous
growth model, 
in  which the population disperses via   diffusion and 
convection is considered. Time-independent environmental heterogeneities,
such as a 
random distribution of nutrients  or  sunlight are modeled 
by  quenched disorder in  the growth  rate.
Linearization of this model of population dynamics shows that 
the fastest growing localized state dominates in a time 
proportional to a power of the logarithm of the system 
size. 
Using an analogy with a Schr$\ddot{o}$dinger
equation 
subject to a constant imaginary vector potential, we propose a
delocalization
transition for the steady state of the nonlinear problem
at a  critical convection threshold separating  localized and 
extended states.
In the limit of high convection velocity,
the linearized growth problem in $d$ dimensions
exhibits singular scaling behavior described by 
a $(d-1)$-dimensional generalization of the noisy Burgers' equation,
with universal singularities in the density of states associated 
with disorder averaged eigenvalues  near the 
band edge in the complex plane.  The Burgers mapping leads to unusual 
transverse spreading of convecting delocalized populations. 
\end{abstract}

\begin{multicols}{2}

\section {Localization and Population Dynamics}

The mathematical analysis of  spatial  patterns in  
biological systems has 
been  an object of intensive research for many years 
\cite{murray,koch}. 
Both the dynamics and the 
equilibrium properties of certain  model  systems have been worked 
out in detail. Biological
processes,  such as the spread of a favored gene, population growth 
of species, ecological competition and so on,  
are often  a  combination of 
diffusion and  convection  with some kind of back  reaction. These 
systems are not conservative;
both growth and death terms, possibly involving nonlinearities, 
change the number of individuals involved.
A general form of such reaction-diffusion equations is \cite{murray}
\be
{\partial \c \over 
\partial t } +(\v\cdot  \nabla)  \c  
=\f (\c) + D \nabla^2  \c,
\eqnum{1.1}
\label{eq:one}
\ee
where $\c (\x , t) $ is the vector of reactants 
(e.g., species of bacteria, nutrients, etc.),
$\D$ is a matrix of diffusivities, and $\f(\c)$ describes   
the nonlinear reaction 
kinetics. The conservative  
term $\v\cdot\nabla\c$ represents a 
convective  flux controlled by a  drift velocity  
$\v$, such as the  flow of water in  aqueous media, winds, etc.      
       
Although the  literature  discussing 
equations of this type is  massive,
the effect of spatial inhomogeneities in the underlying medium 
is relatively unexplored \cite{bose}. A disordered substrate
may manifest itself in the above formalism 
as  quenched random diffusion 
constants,   stochastic  growth and death rates or   randomness 
in the reaction term; 
it may reflect random concentration of environmental factors such as 
nutrients or toxins, or an inhomogeneous illumination
pattern  
projected onto, e.g., photosynthetic bacteria.    

In this paper we study the effect of such  heterogeneities in 
biological  systems. As a model we take one of 
the simplest situations, 
the case of a {\it single}  species, 
described by population number 
density $c(\x,t)$,  
for which the reaction diffusion  equation
is a straightforward  generalization of the Malthus-Verhulst
growth model \cite{murray}:
\begin{eqnarray}
{\partial c(\x,t)\over\partial t} + \v\cdot \nabla c(\x,t) 
&=& D \nabla^2 c(\x,t) + [a + U(\x)]c(\x,t)  \nonumber \\
&& - bc^2(\x,t),
\eqnum{1.2}
\label{eq:two}
\end{eqnarray}
where $U(\x)$ is a zero mean quenched random variable, 
and we take the convective 
velocity $\v$ to be constant in space and time.

The homogeneous analog of this equation, without the convection term
(i.e., $U(\x)=0$ and $\v=0$) 
was proposed by R. A. Fisher \cite{fisher} as a model 
for the spread of
a favorable genetic mutation. 
It  is also useful as a description  of a population dynamics, 
for which
$a$ is the difference between the linear  birth and death rates, 
$D$ reflects the effect of migration, and the term  $-b c^2$ 
represents some 
self-limiting process, roughly proportional to the number of pairs 
of individuals at position $\x$.  In the non-random 
case $U(\x)\equiv 0$,
there are two spatially homogeneous fixed points: an
unstable fixed point at
$c (\x) \equiv {\bf 0}$, in which there is no 
population at all,  and  stable fixed point 
at $c^* (\x)\equiv a/b$,
where the population saturates to the carrying capacity 
of the environment. 
Nonnegative initial  
configurations evolve smoothly toward  the 
stable fixed point; analysis of  the  time development of  
spatial fluctuations  in this model reveals that 
equilibrium can be reached via
traveling solitonlike solutions, known as Fisher waves 
\cite{murray,fisher,kolmo}.

It is interesting to consider (\ref{eq:two}) 
in the context of nucleation 
and spinodal decomposition. In the absence of 
convection, we can rewrite Eq.
(\ref{eq:two}) as 
\be 
{\partial c \over \partial t} = 
- D {\delta F \over \delta c},
\eqnum{1.3}
\label{eq:three}
\ee
where the potential is:

\begin{minipage}[t]{3.2in}
\epsfxsize=3.2in
\epsfbox{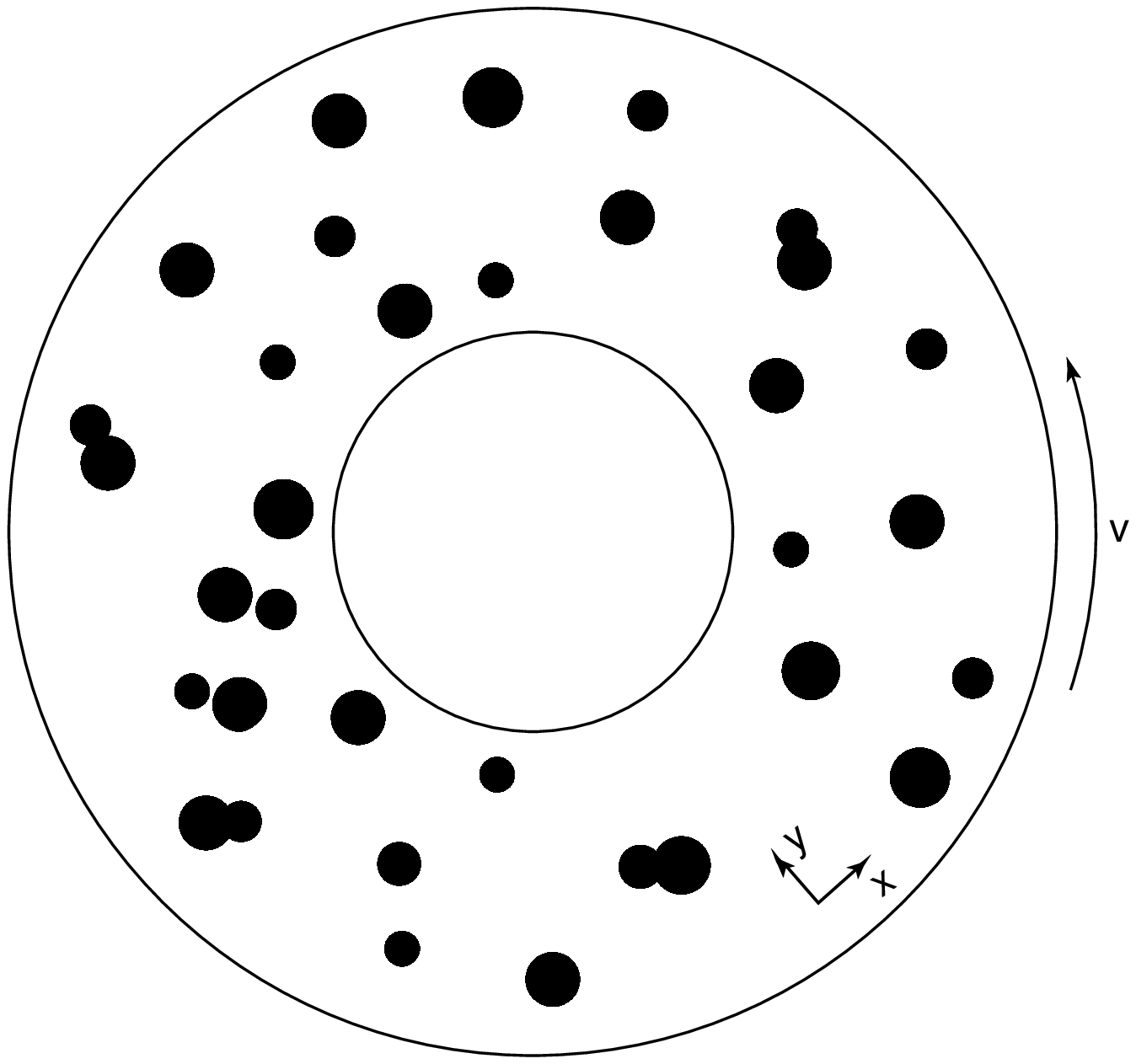}
\begin{small}
FIG.\ 1. 
Schematic of a disordered biological substrate subject
to diffusion and convection. The dark spots represent 
environmental fluctuations, as exemplified by an 
homogeneous pattern of light projected onto a growing 
population of bacteria.
\end{small}
\vspace{0.2in}
\end{minipage}

\be 
F(c)  = {1 \over 2} 
(\nabla c)^2 + \Phi(c)
\eqnum{1.4}
\label{eq:four}
\ee
with
\be 
\Phi(c) = - {1 \over 2D} 
[a + U(\x)]c^2 + {1 \over 3D} b c^3.
\eqnum{1.5}
\label{eq:five}
\ee
The time evolution of a small concentration inhomogeneity thus seems
to resemble  the dynamics of an order parameter quenched 
below its critical 
point and subject to a cubic Landau potential $\Phi(c)$ \cite{domb}.
However, the usual additive Langevin thermal noise 
term is missing in the dynamics.  
This reflects the fact that our system admits 
an absorbing state, i.e., a population state which vanishes
everywhere remains fixed 
at zero.  
A second crucial difference arises because a 
smooth nonnegative initial 
condition will remain nonnegative for all times \cite{proof}. 
The Fisher growth 
model is thus an unusual zero temperature ``one-sided'' spinodal 
decomposition problem. The one-sided nature of the dynamics insures 
that no unphysical effects arise due to the 
unbounded potential (\ref{eq:five}) at large negative $c$.  
Interesting 
studies exist of domain growth in, e.g., random Ising systems at zero 
temperature \cite{li,natter,narayan}, but we are 
unaware of a similar body of work on 
``one-sided'' population growth models. Another 
area  of active research 
concerns zero temperature dynamical systems with multiplicative noise 
\cite {grinstein}. Here, however, the noise typically depends on
both space and time, in contrast to the purely 
space-dependent function 
$U(\x)$ in Eq. (\ref{eq:two}).  
\vskip 8mm

In Eq. (\ref{eq:two}), both  
convection and random fluctuations of the  
growth rate have been added
to the original Fisher model. 
For the sake of concreteness, we can consider Eq. 
(\ref{eq:two})  as a model for  a 
colony of bacteria which grows on an inhomogeneous  substrate. 
The  linear growth  rate of the bacteria depends on  a 
spatially random (but inexhaustible) 
food supply at each point, or  other inhomogeneous, 
time-independent,  
environmental factors such as the intensity of illumination.  
By reducing the light intensity,  this 
linear growth rate could presumably 
assume both positive and negative values 
at different points in space.  
The food or illumination  are quenched random variables. The 
bacteria diffuse, as well as undergo convection due to  flow
of the ambient medium   with velocity $\v$ (Fig. 1)
\cite{note}.

In this simple model of bacterial population growth, we neglect 
the dynamics of a diffusing nutrient supply or feedback
from waste products \cite{murray}. For growth in an inhomogeneous
light source, for example, we might require a fixed homogeneous 
nutrient supply, possibly stabilized by large nutrient diffusion 
constants. Wakita {\it et al.} \cite{wakita} have studied the 
growth of {\it Bacillus subtilis} under various conditions, 
and found a large regime of low Agar
density and high nutrient concentration which is well 
described by the homogeneous analogue of (\ref{eq:two})
without convection, consistent with our 
assumptions.

If the fluctuating growth rate $U(\x)$ is of order $\delta a$, 
we shall see that 
convection will significantly perturb population growth
when the flow rate $\v$  exceeds the corresponding 
change in the Fisher 
wave velocity  \cite{murray} $\delta v_f = 2 \sqrt{D\delta a}$. 
Upon  taking as an effective diffusion constant for a motile 
bacteria $D \sim 6 \times 10^{-6}{\rm  cm}^2/$sec 
\cite{berg} and growth rate fluctuations
$\delta a \sim a  = 
10^{-3}/$sec, we find fluctuating  Fisher wave velocities of at most a few 
microns per second. Thus, we shall  be  
primarily concerned with small values
of $\v$ and low Reynolds number flows in our analysis. 
    
When randomness is introduced into (\ref{eq:two}), constant  
configurations are no
longer  possible steady-state solutions. 
Instead, the steady state is spatially modulated, reflecting
the competition between the 
disorder and the diffusion term.   
Convection, on the other hand, tends to make 
the final state more uniform;
For very high convection velocities, one might 
expect that each bacterium feels
some ``effective medium'' average of the 
random environment as it drifts
rapidly from one site to the other. In this paper, we show 
that there is  a phase transition from one regime to the other 
in the linearized growth problem and suggest that this sharp 
change of behavior persists in the steady state.

In order to simplify our problem, let us assume that a stable 
nonzero steady-state population profile $c^*(\x)$,  
exists, as well as the unstable $c(\x)\equiv 0$
steady state. The time evolution 
of small fluctuations around these (stable or unstable) 
configurations is determined by the   
linearizing Eq. (\ref{eq:two}) near the fixed point
functions.  
Linearizing about the ``Gaussian'' fixed point 
$c(\x) \equiv 0$ leads to
\be
{\partial c\over \partial t} \simeq  D \nabla^2 c -  
\v \cdot \nabla c  
+ [a+  U (\x)] c.
\eqnum{1.6}
\label{eq:six}
\ee
Linearization about the nontrivial 
stable steady-state configuration, 
$c^*(\x)$, which satisfies
\be
D \nabla^2 c^* -  
\v \cdot \nabla c^*  
+ [a + U(\x)]c^*-2bc^{*2} = 0,
\eqnum{1.7}
\label{eq:seven}
\ee
leads to a similar equation for $c'(\x,t) \equiv
c(\x,t)-c^*(\x)$, namely,
\be
{\partial c'\over\partial t}\simeq D\nabla^2 c'-  
\v \cdot\nabla c'  
+ [a+ U'(\x)]c'
\eqnum{1.8}
\label{eq:eight}
\ee
with 
\be
U'(\x) = U(\x)-2bc^*(\x).
\eqnum{1.9}
\label{eq:nine}
\ee
We shall assume  
that $c^*(\x)$ has no long range correlations, so that 
the linear operator in  (\ref{eq:nine}) now involves a redefined  
function $U'(\x)$,  with
quenched random fluctuations away from its mean value 
similar to those described by $U(\x)$.

One main interest in this paper is the evolution of small
fluctuations about these  fixed point  
configurations. Near the unstable 
$c(\x) \equiv 0$ 
configuration, fluctuations grow until they reach the point 
for which the linearization fails. On the other 
hand, small perturbations of 
the stable state $c^*(\x)$ will decay, so 
that the linear approximation 
becomes 
better in time. Thus,  Eqs. (\ref{eq:six}) and (\ref{eq:eight}),
describe the short time growth
of dilute populations or the long time decay 
to the stable state. Although 
we shall frame most of our discussion in terms of the unstable modes 
described by (\ref{eq:six}), a very similar analysis applies to Eq. 
(\ref{eq:eight}). 

Similar equations emerge in a variety of physical situations,
such as models of chemical reactions and neural 
network \cite{murray}. 
Miller and Wang {\it et al.} \cite{miller} have 
recently studied the spectrum of an operator describing 
diffusion of a passive scalar subject to a spatially random, 
time-independent velocity field, but without simple multiplicative
randomness.
A closely  related physical system from which we intend  to draw, 
is fluctuating vortex lines in superconductors 
in  the  presence of columnar defects,  
with external magnetic field tilted away 
from the direction parallel to the 
defects \cite{nel-dous,nel-vin}. 
The partition function $\cZ(\x,t)$  for a single line
at position $\x$ and height $t$  
then satisfies Eq. (\ref{eq:six}) where $D$ is given by   
temperature divided by the tilt modulus,
$D = T/(2\teps_1)$, $\v $ is proportional  to the 
tilt field, and $U(\x)$ corresponds
to  the columnar disorder potential
in the superconductor $V(\x)$ 
normalized by the temperature, $U(\x) = V(\x)/T$. 
Some basic facts about 
vortex lines are revived  in Appendix A.
It is interesting to note that Eq. (\ref{eq:six}) also 
describes the growth of monetary capital with diffusion, spatially 
varying interest rates and drift due to, say population 
migration. 

Another related problem concerns diffusion and drift of particles 
in a medium with randomly distributed traps \cite{haarer,movaghar,grassberger}.
The long time decay of the density of active particles as obtained 
experimentally from,   e.g.,  photoconduction studies 
in quasi one dimensional polymers \cite{haarer}, is expected to 
exhibit stretched exponential relaxation in the absence of a bias 
\cite{movaghar,grassberger}. When a biasing electric field 
is present, the decay has a simple exponential prefactor \cite{grassberger}
with however a transition from stretched exponential to exponential 
decay in the subleading behavior above a critical bias threshold 
\cite{movaghar}. The coarse grained physics can be approximated by 
Eq. (1.2) without the nonlinearity and $U(\x)$ chosen so that {\it all}
growth eigenvalues are negative when ${\bf v} = 0$. Our primary concern 
here is with situations where at least some growth eigenvalues 
are positive. However, a delocalization transition has also been 
invoked to describe an abrupt onset of a drift velocity as a function 
of the bias as the particle density decays in the trapping problem 
\cite{grassberger}.  We discuss the relation of the results of 
Ref. \cite{movaghar} and \cite{grassberger} to the population biology problem 
treated here in Appendix C.

After a  Cole-Hopf transformation, i.e.,
\be 
c(\x,t) = \exp[\lambda\Phi(\x,t)] ,
\eqnum{1.10}
\label{eq:ten}
\ee
the linear growth model described by Eq. (\ref{eq:six}) becomes
\begin{eqnarray}
\partial_t \Phi(\x,t) &=& 
D\nabla^2 \Phi(\x,t) + 
\lambda D ({\bf \nabla} \Phi)^2 \nonumber \\ 
&-&\v \cdot {\bf \nabla} \Phi(\x,t) +  
a +  U(\x),
\eqnum{1.11}
\label{eq:eleven}
\end{eqnarray}
while $c'(\x,t) = \exp[\lambda\Phi'(\x,t)]$ generates
a similar equation from Eq. (\ref{eq:eight}) with $U(\x) \to U'(\x)$.
Chen {\it et al.} \cite{Chen} have proposed Eq. (\ref{eq:eleven}) 
as a model for the 
dynamics of strongly 
driven  charge-density waves  
with quenched disorder. 
Later in this paper, we use Eq. (\ref{eq:eleven}) to 
study sample-to-sample fluctuations in $ln c(\x,t)$ in the 
limit of high convection velocities. We obtain exact analytic 
results, which should be applicable both to linearized models
of population dynamics and to charge
density waves. 
 
Equations (\ref {eq:six}) and (\ref{eq:eight}) 
may also  be written as  
\be
\partial_t c=\cL c,
\eqnum{1.12}
\label{eq:twelve}
\ee
where the Liouville operator, e.g., 
\be
\cL = D\nabla^2-\v\cdot{\bf \nabla} +a + U(\x),
\eqnum{1.13}
\label{eq:thirteen}
\ee
generates the time evolution of the system. The spectra 
and eigenvalues of random non-Hermitian operators similar
to Eq. (\ref{eq:thirteen}) have attracted considerable 
interest recently \cite{janik}.
Provided linearization is an adequate approximation, 
the dynamics of this system is
determined by the eigenvalues and the eigenvectors of $\cL$.
Near the stable fixed point $c^*(\x)$  one  expects only 
decaying modes, i.e., all
real parts of the 
eigenvalue spectrum of  $\cL$ are  negative, while near the unstable 
state $c(\x)\equiv 0$ 
there are at least few positive, growing eigenstates. 
These expectations  can be demonstrated 
explicitly when randomness is absent,
i.e., $U(\x) = 0$. In this case the right 
eigenvectors of the non-Hermitian operator (\ref{eq:thirteen})
about the unstable fixed point are simple plane 
waves $\phi_\k^R(\x) \sim
e^{i\k \cdot \x}$ which satisfy
\be 
\cL \phi_\k^R(\x) = \Gamma_\k \phi_\k^R(\x),
\eqnum{1.14}
\label{eq:fourteen}
\ee
with the complex eigenvalue spectrum
\be 
\Gamma(\k)=a-i\v\cdot\k-Dk^2.
\eqnum{1.15}
\label{eq:fifteen}
\ee
The operator $\cL$ which corresponds to linearization about the 
nontrivial fixed point $c^*(\x)=a/b$ has the same eigenfunctions,
with spectrum
\be
\Gamma'(\k)=-a-i\v\cdot\k-D k^2.
\eqnum{1.16}
\label{eq:sixteen}
\ee
Provided $a>0$, the eigenvalues
of (\ref{eq:fifteen}) have a positive real 
part for small $\k$, while all
the eigenvalues of (\ref{eq:sixteen}) have a negative real part.
Note that the eigenfunctions are always delocalized plane
waves.

It is instructive to use the full 
nonlinear equation (\ref{eq:two}) (with $U(\x)=0$) to trace
the time evolution of a small random initial condition 
$c(\x,t=0)$ in terms of the above eigenmodes when $a>0$. Assume 
that initial conditions have contributions at all wave vectors 
$\k$ including $\k = 0$. Although {\it many} eigenmodes are 
unstable near $c(\x) \equiv 0$ and grow exponentially in time, 
the $\k=0$ mode grows most rapidly. Once this uniform mode 
saturates at its fixed point value, it then acts 
to suppress all other growing modes via mode coupling induced
by the nonlinear mode coupling term $-bc^2(\x,t)$. These expectations 
are illustrated via a mean field solution of Eq. (\ref{eq:two}) in 
Appendix B.

The spatial characteristics of the eigenfunctions 
of the linearized growth operator 
change dramatically when randomness is present.
When  $\v = 0$, the  operator  
$\cL=D \nabla^2+a+U(\x)$ 
is Hermitian with real eigenvalues, and for strong 
enough disorder, all its eigenfunctions 
are real and localized; the 
localization length is smallest in the tails 
of the energy band, corresponding 
to extreme  values of $U(\x)$. In one and two dimensions, 
it is widely
believed that {\it all} 
states are localized even for weak disorder \cite{sh-ef}. 

When $\v \neq 0$, the Liouville growth operator is no 
longer Hermitian, although 
it  can still  be diagonlized using a system of  
left and right eigenvectors. 
Since the  convection  
term in Eq. (\ref{eq:six}) 
may be absorbed into the Laplacian by completing the square, 
${\bf \nabla} 
\to {\bf \nabla}+{\v \over 2D}$, the  right and left eigenfunctions
of the new Liouville operator  
are related to the eigenfunctions of $\cL(\v=0)$ via an imaginary 
``gauge transformation'';
if $\phi_{n,\v=0} (\x)$ is an eigenfunction of the 
Hermitian problem, then 
\cite{fokker}
\begin{eqnarray}
\phi_{n;\v}^R(\x)&=&e^{\v \cdot \x/D} \phi_{n;\v = 0}(x)  
\nonumber \\  
\phi_{n;\v}^L(\x)&=&e^{-\v \cdot \x / D} \phi_{n;\v = 0}(x)
\eqnum{1.17}
\label{eq:seventeen}
\end{eqnarray}
are the  eigenfunctions of the non-Hermitian 
operator with the {\it same}
eigenvalue $\epsilon_n$, up to a constant shift \cite{grassberger}, 
\be
\epsilon_n \to   \epsilon_n ({\bf v}) = \epsilon_n ({\bf v}=0) - {v^2 \over 
4D} \nonumber 
\ee
provided that $\xi_n$,  the localization 
length in the non-driven problem, is less then $D/\v$. 
Thus, for small 
convection velocities, 
there is a {\it spectral rigidity} - except for the shift, 
 the real  eigenvalue spectrum 
is  locked to the values it had for $\v=0$. 
As  $\v$ increases, however, some  eigenfunctions  eventually  become 
extended and 
the dynamics becomes  sensitive to   
boundary effects.  
The
correct eigenfunctions end eigenvalues  are no longer  
related  to the $\v=0$ case by a simple
transformation. In case of 
periodic boundary conditions, complex eigenvalues
and delocalized modes  
appear when $D/v$ becomes 
smaller than $\xi_n$. As $\v$ is  increased, 
these delocalized states appear
first at the band center, for which the 
localization length is maximal,  
then move outwards. 

These expectations for the eigenvalue spectrum   
have been demonstrated  by analytic work and 
a numerical analysis of a discrete 
lattice model, inspired by the physics of vortex lines \cite{hat}.
The corresponding lattice 
discretization of the nonlinear equation which 
motivates our present work
reads
\begin{eqnarray}
{dc_\x(t) \over dt}& =& {w \over 2}
\sum_\x \sum_{\nu = 1}^d 
[e^{\g \cdot \e_\nu}c_{\x +\e_\nu}(t) +
e^{-\g\cdot\e_\nu} c_{\x-\e_\nu}(t)]
\nonumber \\
&&+ \sum_\x 
[a + U(\x)]c_\x(t)-b\sum_\x c_{\x}^2(t)\;,
\eqnum{1.18}
\label{eq:eighteen}
\end{eqnarray}
where $c_\x(t)$ is the species 
population at the sites $\{\x\}$ of a hypercubic 
lattice, and the  $\{\e_\nu\}$ are unit lattice vectors. Here $w$
is proportional to the diffusion constant of the 
corresponding continuum model $(w \sim D/\ell_0^2)$, 
and $\g$ is proportional to the flow rate
$\v$. Up to subtraction proportional to $v^2/D$  in $a$, $a$, $U(\x)$ and $b$ 
have the same interpretation as in the continuum
model. When linearized about $c_{\x}  \equiv 0$, 
Eq. (\ref{eq:eighteen}) may be written 
\be 
{dc_\x (t) \over dt} = \sum_{\x'}{\tilde \cL}(\x,\x') 
c_{\x'}(t)
\eqnum{1.19}
\label{eq:nineteen}
\ee
where the discrete Liouville operator ${\tilde\cL}$ is the matrix 
\begin{eqnarray} 
\tilde \cL&=&{w\over 2}\sum_\x \sum_{\nu=1}^d 
[e^{-\g\cdot\e_\nu}|\x + \e_\nu\rangle \langle\x| \;  + \;
e^{\g\cdot\e_\nu}|\x\rangle \langle\x +\e_\nu | ]  \nonumber \\
&&+
\sum_\x [a+U(\x)] |\x\rangle\langle\x|
\eqnum{1.20}
\label{eq:twenty}
\end{eqnarray}
As in the continuum case, the same equation 
arises when linearized about
the a nontrivial fixed point $c^*_\x$ provided we make the 
replacement $U(\x)\to U'(\x)$ with $U'(\x)$ given 
by Eq. (\ref{eq:nine}) 

Typical spectra for a 1000-site model in one dimension with $U(\x)$
uniformly distributed in the interval 
$[-\Delta,\Delta]$ are shown in Fig. 2 for three
values of $g \propto v$ \cite{hat}. 

\begin{minipage}[t]{3.1in}

\epsfxsize=3.1in
\begin{small}
(a)
\end{small}

\epsfxsize=3.1in
\epsfbox{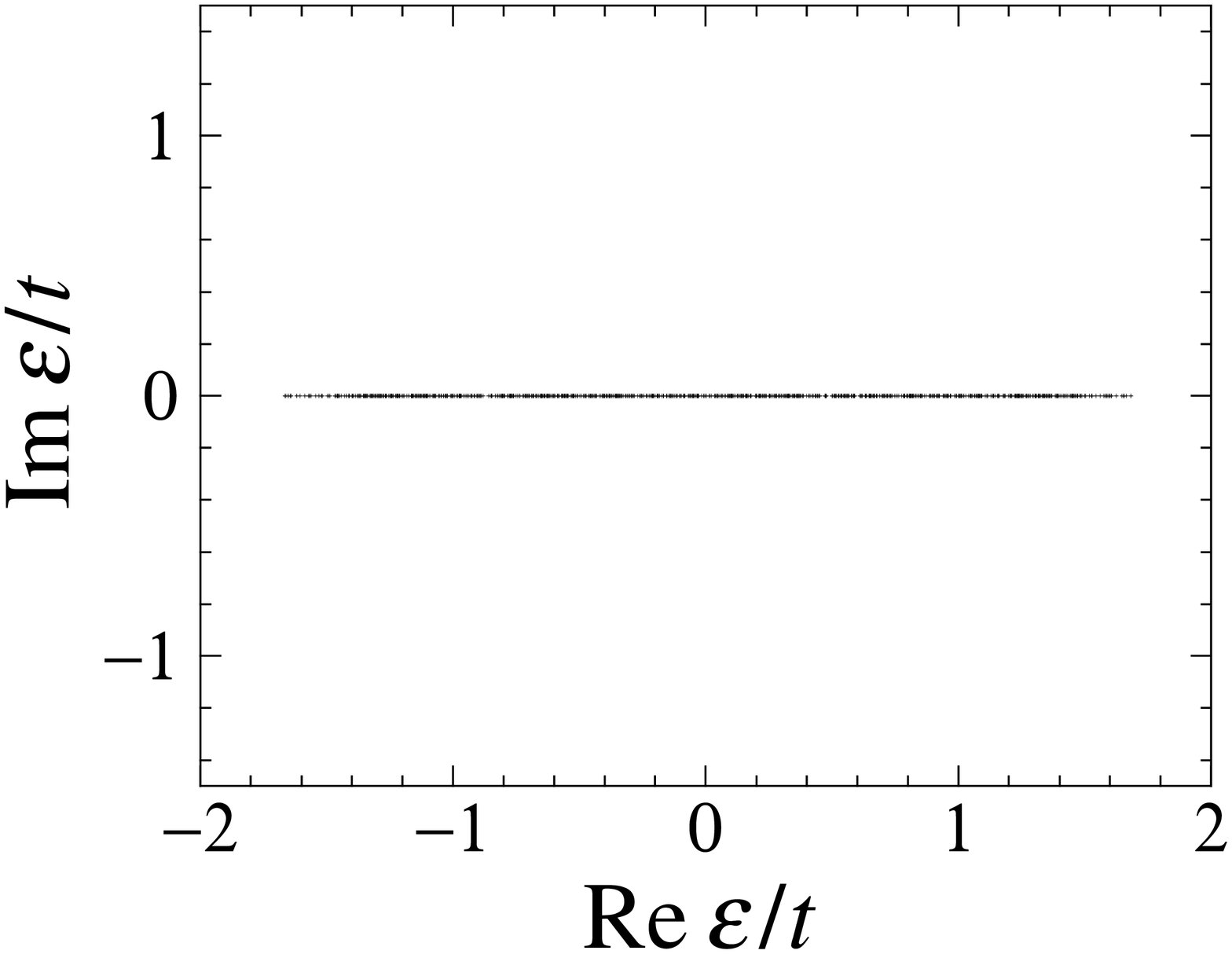}
\end{minipage}

\begin{minipage}[t]{3.1in}
\epsfxsize=3.1in
\begin{small}
(b)
\end{small}

\epsfxsize=3.1in
\epsfbox{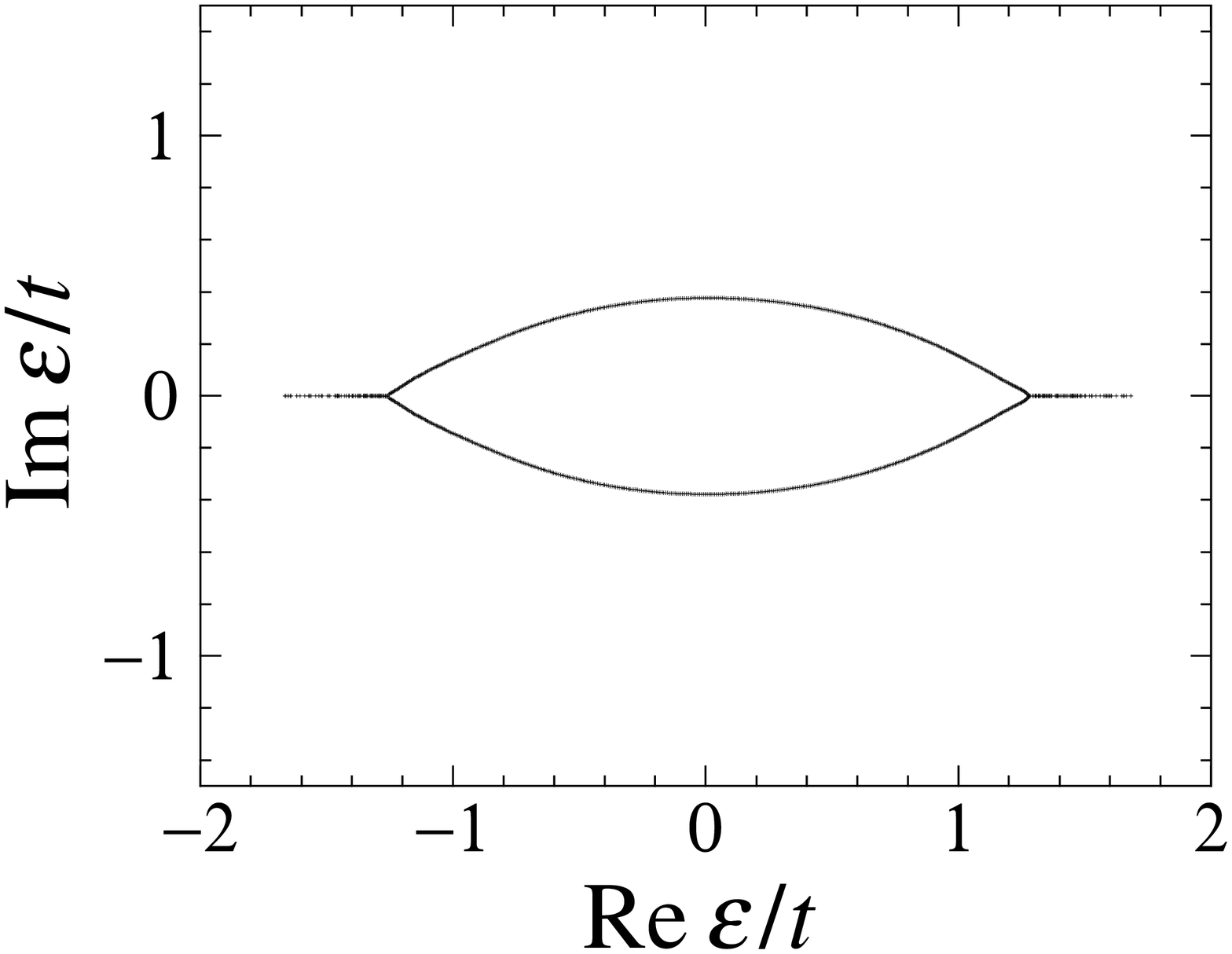}
\end{minipage}

\begin{minipage}[t]{3.1in}
\epsfxsize=3.1in
\begin{small}
(c)
\end{small}

\epsfxsize=3.1in
\epsfbox{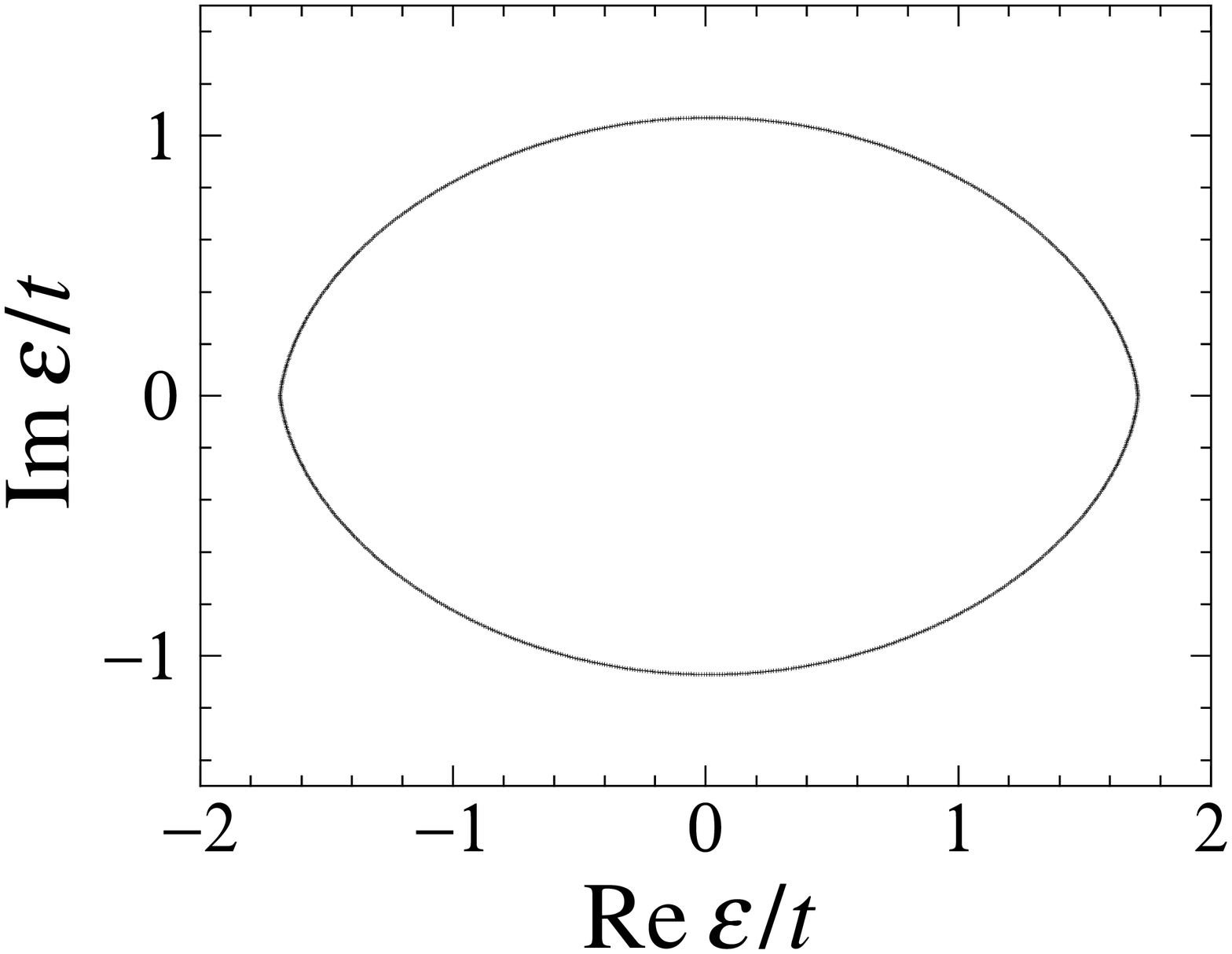}
\begin{small}
FIG.\ 2. 
Energy spectra of one-dimensional 1000-site lattice
model with randomness $\Delta/w=1$. The resulting 
spectrum for the same realization of the random potential
$U(x)\in[-\Delta,\Delta]$ is plotted here for three 
different values of $g$. (a) Case $g<g_1$; all eigenstates 
are localized; (b) $g_1<g<g_2$; bubble of complex 
eigenvalues indicating extended states appears near the center 
of the band; (c) $g_2<g$; all the eigenstates are 
extended. (After Ref. \cite{hat}).
\end{small}
\vspace{0.0in}
\end{minipage}

\noindent
For $g$ less 
than a critical value $g_1$,
all eigenmodes are localized, and 
the eigenvalues remain real and locked to their  values
for $g=0$. For $g_1<g<g_2$, extended states with complex eigenvalues 
appear near the center of the band. Localized states still appear
near the band edges. For $g>g_2$, every 
localized state is destroyed by 
the non-Hermitian perturbation, and all states are extended.     
In this limit, eigenfunctions are slightly 
perturbed Bloch states---the lattice version of plane waves.
The spectrum is well approximated by the disorder-free limit,
i.e., the lattice analogue of Eq. (\ref{eq:fifteen}) 
\be 
\Gamma(k)=2w\cos(k\ell_0 + ig\ell_0),
\eqnum{1.21}
\label{eq:tone}
\ee
where $\ell_0$ is  the lattice constant.

With our definition of $\tilde {\cL}$,
states near the {\it top} of the band should give a 
reasonable approximation to 
the spectrum of the  continuous operator 
(\ref{eq:fourteen}); Eq. (\ref{eq:fifteen})
then describes the upper edge of the 
ellipse of eigenvalues in Fig. 2c.
The  states at the bottom 
of the band, on the other hand, 
have spatial characteristics which are 
artifacts of the lattice discretization.

Figure 3 shows typical spectra for the 
discrete operator in {\it two} dimensions
\cite{hat}. Here, too, eigenfunctions near the 
top of the band should give
good approximation to growth modes in the continuum limit.
Again, all eigenvalues remain real and the eigenfunctions remain  
localized    
when $g<g_1$. For $g_1 < g < g_2$, 
however, extended and localized states 
{\it coexist} near the center of the band 
\cite{hat}. When $g>g_2$,  
even most rapidly growing states at the top of the 
band are delocalized.
For {\it very} large $g$ (not shown) the 
spectrum resembles that of the 
disorder-free limit of the lattice model, 
similar to the one-dimensional case.
However, as mentioned in Ref. \cite{hat}, this 
apparent simplification is 
actually a finite size artifact in $d=2$: 
level repulsion leads to large 
modification of the Bloch wave functions and 
eigenvalues even for weak disorder 
in any sufficiently large system. Thus, chaotic spectra similar 
to Fig. 3c are always expected for large $v$ due to disorder 
in the ``thermodynamic
limit'' of large system sizes in $d=2$. In Sec. 4 we argue 
that the growth for $g>g_2$ when $d \geq
2$ is in fact described by the $(d-1)$-dimensional 
Schr\"odinger-like equation with space- and 
time-dependent randomness. The anomalous critical 
exponents which describe this situation lead to 
universal power law  singularities in the density of states near 
the band edge in the complex plane.

If we take a gradient and set $\u ( \x , t)=- \nabla
\Phi(\x,t)$, the dynamical growth model (\ref{eq:eleven}) 
which results from a Cole-Hopf transformation reads
\begin{eqnarray} 
\partial_t \u ( \x , t)&+&(\v\cdot\nabla)\u ( \x , t)+2\lambda D
(\u(\x,t)\cdot\nabla)\u( \x , t) \nonumber  \\
&=&
D\nabla^2\u( \x , t)+{\bf f}(\x),
\eqnum{1.22}
\label{eq:twentytwo}
\end{eqnarray}
with ${\bf f}(\x)={\bf \nabla}U(\x)$ and subject to the constraint 
${\bf \nabla}\times\u=0$. This is a variant of the $d$-dimensional 
generalization of Burger's equation with noise studied 20 years
ago by Forster {\it et al.} \cite{Forster}. In the form 
(\ref{eq:eleven}), such problems are sometimes referred to 
as ``KPZ equations''.

\begin{minipage}[t]{3.in}
\epsfxsize=3.in
\begin{small}
(a)
\end{small}

\epsfxsize=3.in
\epsfbox{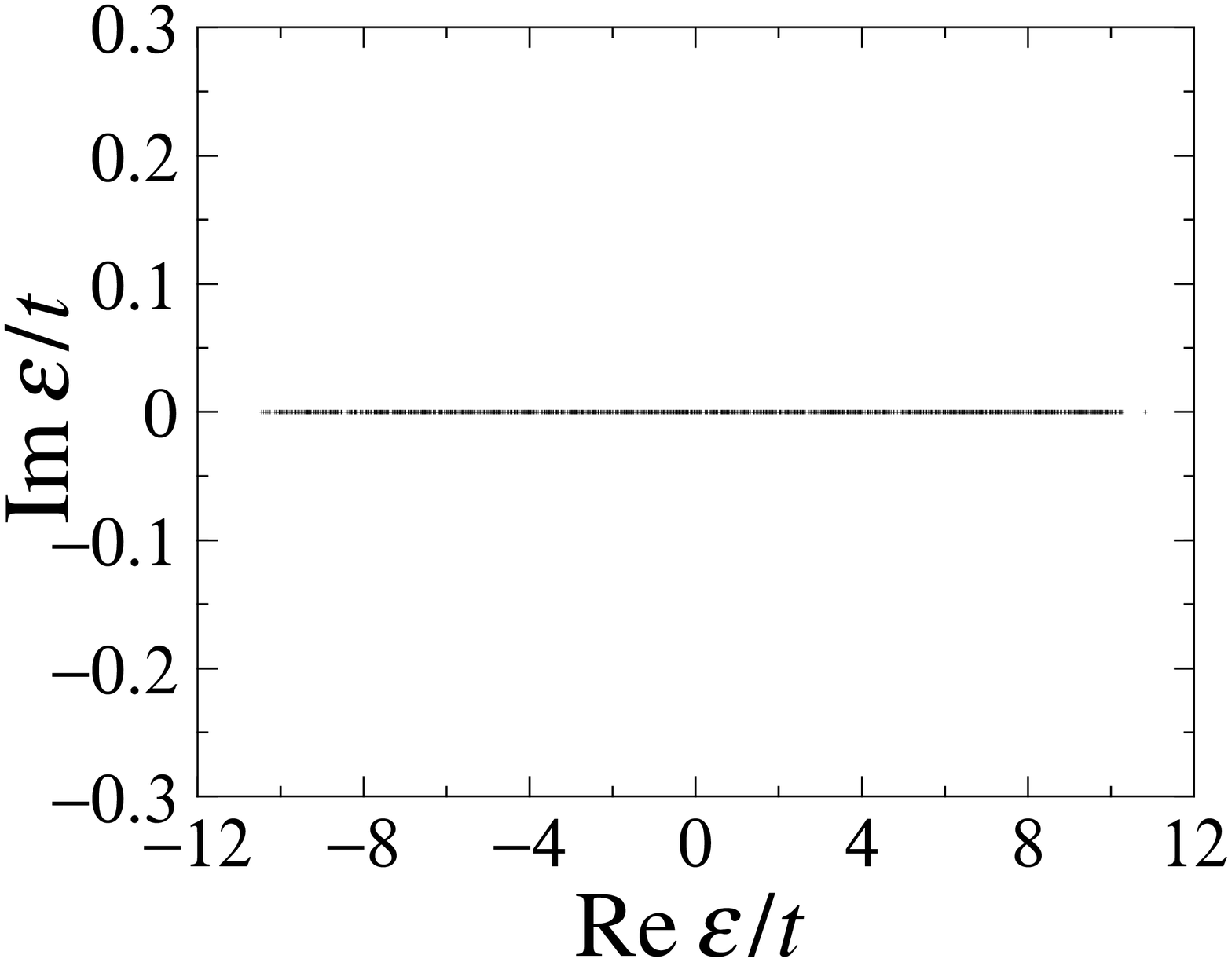}
\end{minipage}

\begin{minipage}[t]{3.in}
\epsfxsize=3.in
\begin{small}
(b)
\end{small}

\epsfxsize=3.in
\epsfbox{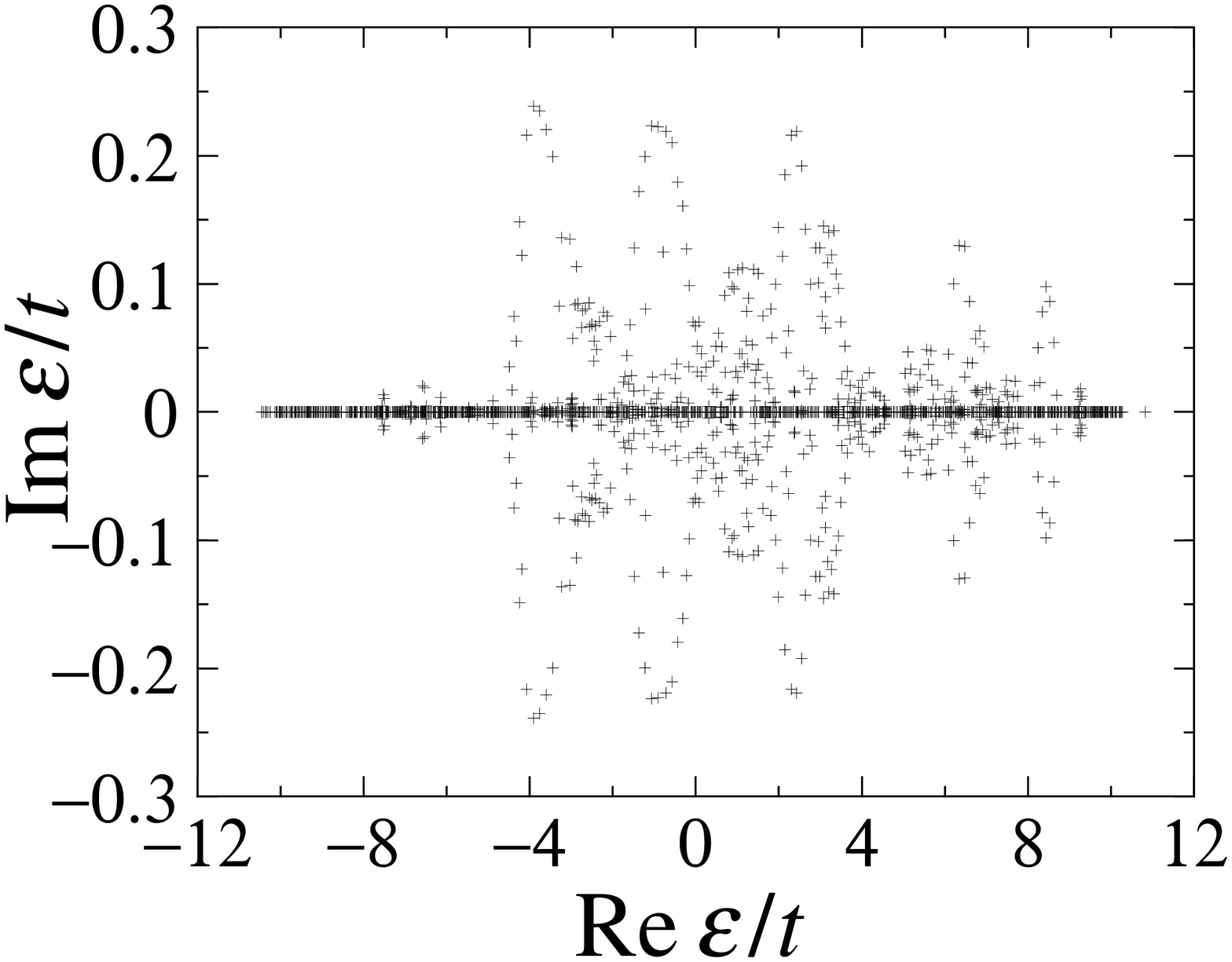}
\end{minipage}

\begin{minipage}[t]{3.in}
\epsfxsize=3.in
\begin{small}
(c)
\end{small}

\epsfxsize=3.in
\epsfbox{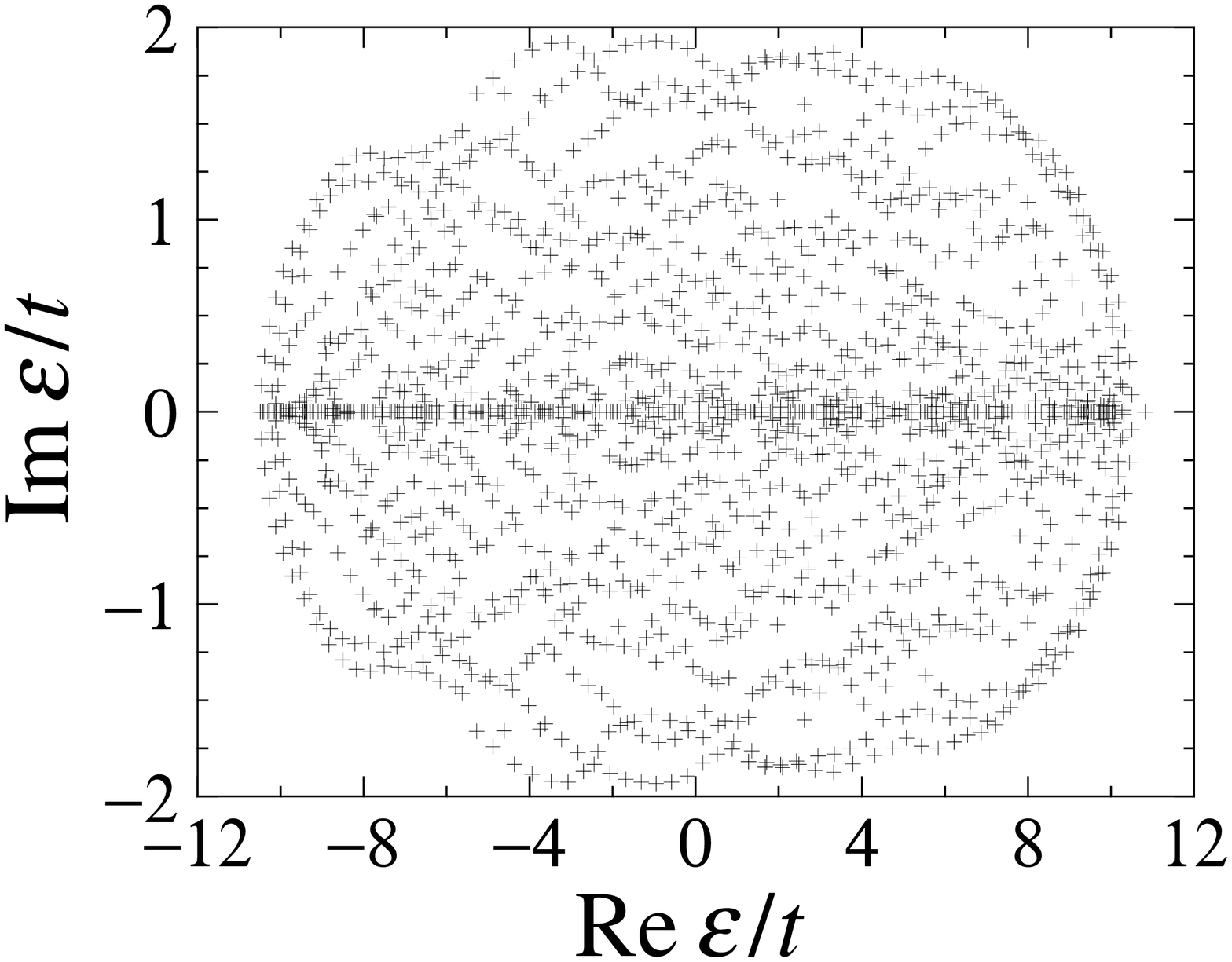}
\begin{small}
FIG.\ 3. 
Typical spectra of the two-dimensional tight-binding
non-Hermitian model with random site potential. (a) Case 
$g<g_1$; delocalized states, real spectrum. (b) $g_1<g<g_2$; 
extended states with complex eigenvalues coexist with localized 
states with real eigenvalues near the center of the band. 
(c) $g_2<g$; states at the tails of the band become 
extended, and there are complex eigenvalues even near the top 
of the band. (After Ref. \cite{hat}.)
\end{small}
\vspace{0.0in}
\end{minipage}

\noindent
In Sec. 5, we show that (\ref{eq:twentytwo}) is in the same 
universality class as a simpler $(d-1)$-dimensional 
noisy Burgers' equation, and use this fact 
to study sample-to-sample fluctuations of 
$ln [c(\x,t)]$ for large $v$. 
Thus, as the convective velocity 
$\v$ grows from small toward large values, 
the most rapidly growing modes
evolve from a dynamics described by Anderson localization
into a regime related to the Burgers' model 
for turbulence \cite{burgers}.

The remainder of this paper is organized as 
follows. In Sec. 2, we assume 
convective effects are small, and illustrate the consequences 
of localization for ``unbounded'' population growth, i.e., 
growth at times before  the nonlinear term in Eq. (\ref{eq:two})      
becomes important. In Sec. 3, we describe how these nonlinearities 
affect the growth when only a few modes are unstable 
relative to the state $c(\x) \equiv 0$. 
{\it Unlike} growth in a homogeneous 
environment, the most rapidly growing 
eigenfunction does not suppress 
all other unstable modes in this case. 
We suggest that the steady state undergoes  a  delocalization 
transition which occurs when the average 
growth rate or the convective 
velocity $\v$ are increased.

In Sec. 4, we 
study the linearized growth problem 
in the large $v$ limit,  show that the 
physics is related to a $(d-1)$-dimensional 
Schr\"odinger equation and demonstrate  that the average 
growth spectrum is singular for $d\geq 2$. This mapping 
leads to {\it universal} predictions for the randomness-dominated 
transverse wandering of a delocalized population as it 
drifts  downstream.   A $d-1$-dimensional 
Burgers' equation is used to describe sample to sample  fluctuations 
in $ln [c(\x,t)]$ in the limit of large $\v$ 
in Section  5.
A number of related calculations are contained in three 
Appendices. Appendix C contains a discussion of the stretched 
exponential relaxation expect for populations in a medium with randomly 
distributed traps. 

\section{Unbounded Growth and the Localized Limit}

There is large literature on localization of 
electrons in disordered 
semiconductors \cite{sh-ef}. Because the 
linearized growth modes which arise 
from Eq. (\ref{eq:two}) obey a Schr\"odinger-like 
equation, one might 
hope that results from electron localization 
theory would be applicable to the simple 
model of heterogeneous population 
dynamics discussed above. Of course, electronic  
states are either empty or full, 
according to Pauli exclusion principle. In contrast, {\it 
many} individual members of a species participate 
in the growth modes determined
by the continuum model equation studied here. 
Another difference 
is the form of the growth-limiting 
nonlinearity in Eq. 
(\ref{eq:two}).

In this section, we explore the consequences 
of localization for the simple problem of 
``unbounded growth'', i.e., growth at times before  the 
nonlinearity in Eq. (\ref{eq:two}) becomes important. 
We assume homogeneous initial conditions, 
weak convection, and parameters (small $b$, for example) 
such that the time domain over which unbounded growth occurs is 
very large.
Effects of the nonlinear term will be discussed in Sec. 3. 

Although many modes may now be growing 
exponentially, the fastest growing 
eigenfunction eventually dominates the center of mass of the evolving
 population distribution. We show here 
that the time $t^*$ 
it takes for the
``ground state'' (i.e., the {\it fastest} 
growing eigenfunction) to win 
out in a large but finite domain grows very slowly as a
function of  
the domain radius $R$.
The precise form of $t^* (R)$
depends on the  
behavior  of the density of localized states in the tail of the band 
of growth eigenvalues. 
For the simple discretized growth model 
discussed in Sec. 1, $t^* (R) \sim 
\ln^{2/d}(R)$ 
where $d$ is the dimensionality of  space. 
Comparable results in semiconductors are usually 
determined by electrons at Fermi energy in 
a partially filled band, and hence 
are less sensitive to the form of the 
density of states. The time required 
for ground state dominance when populations grow 
in a spatially  homogeneous environment is very different, 
$t^*(R)\sim R^2$.  

Given a time-independent Liouville operator $\cL$, as in Eq. 
(\ref{eq:thirteen}), we describe 
growth in terms of a complete set of  
left and right eigenvectors $\{\phi_n^L(\x)\}$ 
and $\{\phi_n^R(\x)\}$,
with eigenvalues $\{\Gamma_n\}$ (the complex ``energy spectrum'').
The time evolution of $c(\x,t)$ is then given by
\be
c(\x,t) = \sum_n c_n \phi_n^R(\x) e^{\Gamma_n t}
\eqnum{2.1}
\label{eq:twoone}
\ee
where the coefficients $\{c_n\}$ are 
determined by the initial condition,
\be
c_n = \int d^d\x\ \phi_n^L(\x) c(\x,t=0)
\eqnum{2.2}
\label{eq:twotwo}
\ee

At long times the system will be dominated by the ``ground state'',
i.e., the state for which the real part $\Gamma_n$ is {\it maximal}. 
Throughout this paper, we assume that there is 
such a state, i.e., that the real part 
of the spectrum is bounded above.  The Liouville operator in
Eq. (\ref{eq:thirteen}) plays, with 
the replacement $t \to - i t$ and $\cL \to -\cH$
a role similar to the Hamiltonian $\cH$ in 
the Schr\"odinger 
equation $i \hbar \partial_t c=\cH c$. 
We shall often use nomenclature from quantum 
mechanics, such as ``energy spectrum'' and ``low-lying states''.
However, because of the identification $\cH\equiv-\cL$, the ground state
of the Hamiltonian is actually the state with the {\it maximal}
eigenvalue of the Liouville operator, and the low-lying states 
of the Hamiltonian are  those which grow fastest for the Liouville 
operator. 

We assume a small convective velocity, 
so that the left and right
eigenfunctions are described by slightly distorted 
versions of the localized 
state for $\v = {\bf 0}$, as in 
Eq. (\ref{eq:seventeen}), with normalization
\be
\int d^d\x \ \phi_m^L(\x)\phi_n^R(\x)=\delta_{m,n}.
\eqnum{2.3}
\label{eq:twothree}
\ee
Deep in the band tail and close to the ground 
state, the $n^{{\rm th}}$ localized 
eigenfunction for $\v={\bf 0}$ will have the approximate form
\be
\phi_{n,\v=0}(\x) \sim b_n e^{-\kappa_n|\x-\x_n|},
\eqnum{2.4}
\label{eq:twofour}
\ee
where $b_n \propto \kappa_n^{d/2}$ is a normalization constant and 
$\kappa_n$ is the inverse localization length associated with 
an eigenmode located at position $\x_n$.  

Of course, only the ground 
state is guaranteed to be strictly nonnegative 
\cite{perron}, as implied 
by Eq. (\ref{eq:twofour}). Orthogonality with the ground 
state requires a small negative part in the localized
excited state eigenfunctions. Nevertheless, Eq. 
(\ref{eq:twofour}) should be a good 
approximation

\begin{minipage}[t]{3.2in}
\epsfxsize=3.2in
\epsfbox{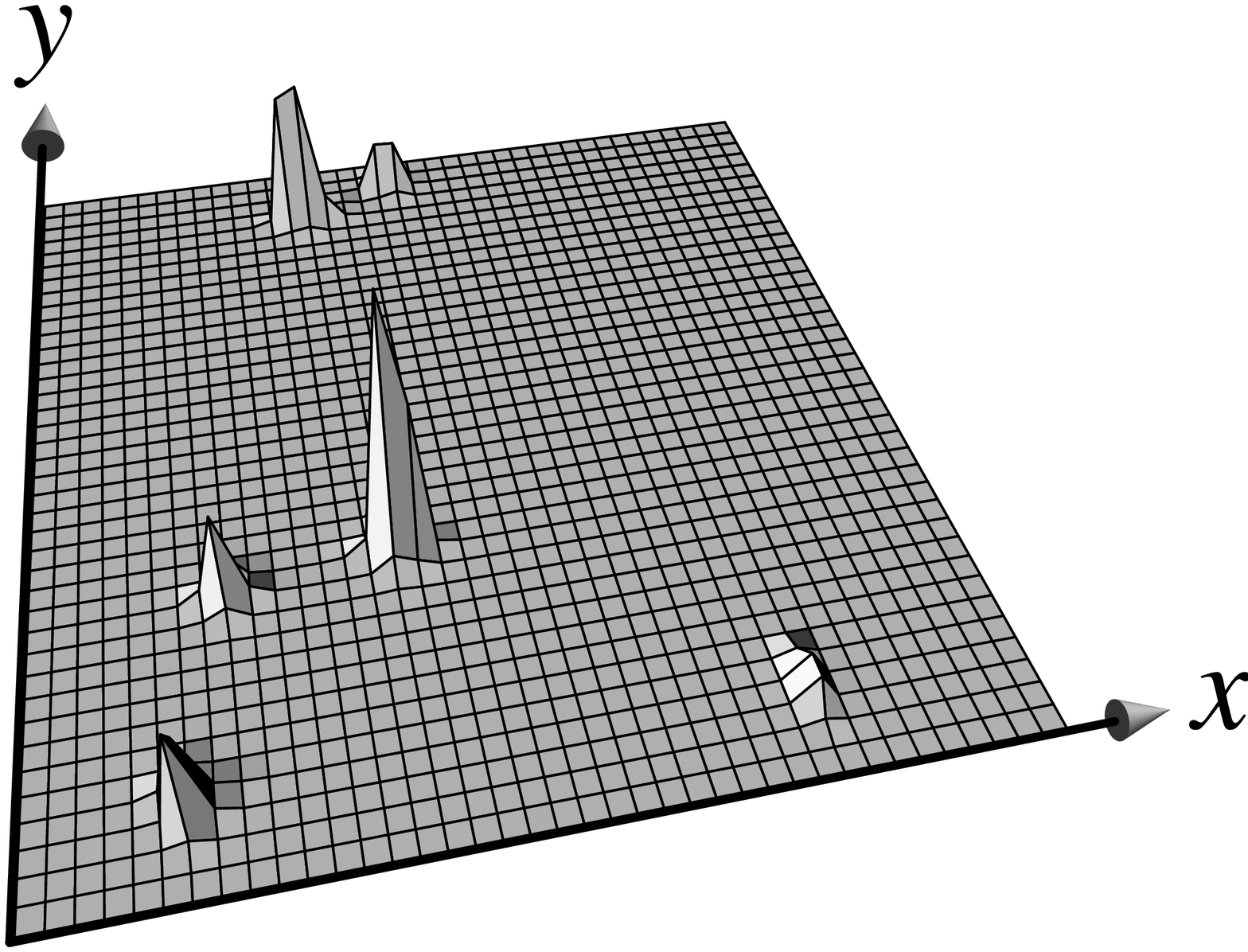}
\begin{small}
FIG.\ 4. 
The six fastest growing 
localized eigenfunctions for the two-dimensional lattice 
model (with $v\propto g=0$) which gave rise to the 
spectrum of Fig. 3(a).(Figure courtesy of Naomichi 
Hatano)
\end{small}
\vspace{0.25in}
\end{minipage}

\noindent
for all modes in the tail of the growth spectrum,
provided the disorder is strong: 
Strong disorder means that all states are well localized, with 
approximately 
the same localization length. The six 
fastest growing modes on the square 
lattice which gave rise to the spectrum in 
Fig. 3a are shown in Fig. 4. 

Suppose the environment has been inoculated with 
a small  {\it uniform}
population of, say, bacteria at time $t=0$. 
Then the projections $\{c_n\}$
of this initial condition onto the localized eigenmodes 
are approximately equal,
$c_n  \approx c_0$ for all $n$. The 
normalized species concentration, 
\be
\hat c(\x,t) ={ c(\x,t)\over\int d^d\x \;  c(\x,t)}
\eqnum{2.5}
\label{eq:twofive}
\ee
is given by
\be
\hat c(\x,t) \approx{\phi_{gs}^R(\x) +\sum_n'\phi_n^R(\x) 
e^{-(\Gamma_{gs}-\Gamma_n)t}\over  N_{gs} + 
\sum_n' N_ne^{-(\Gamma_{gs}-\Gamma_n)t}}
\eqnum{2.6}
\label{eq:twosix}
\ee
where $N_{gs}$ and the $\{N_n\}$ are normalization constants,
\be
N_{gs} = \int d^d\x \;\phi_{gs}^R(\x), 
\qquad 
N_{n}=\int d^d\x \; \phi_{n}^R(\x).
\eqnum{2.7}
\label{eq:twoseven}
\ee 
We have separated out the ``ground state,'' i.e., the most 
rapidly growing mode
and the sums $\sum_n'$ are over the remaining excited states. 

The time necessary for the ground state to dominate the normalized 
species concentration in a large region of 
size $R$ is clearly of order 
$t^*  \sim 1/(\Gamma_{gs}-\Gamma_n^*)\equiv 1/\Delta\Gamma$,
where $\Gamma_n^*$ is the growth rate of the 
first excited state in the 
region. Let $g(\Gamma)$ be the density of states per unit volume 
with growth rates  in the interval between $\Gamma$ and 
$\Gamma + d\Gamma$. Then, as we 
increase $\Delta \Gamma$ from zero, an excited 
state with gap $\Delta \Gamma$ will appear when
\be 
R^d g(\Delta\Gamma)\Delta\Gamma\approx 1 .
\eqnum{2.8}
\label{eq:twoeight}
\ee  
If $g(\Delta \Gamma)$ were approximately constant near the 
band edge, then $\Delta \Gamma \sim 1/R^d$ and we would have
$t^*(R) \sim R^d$. However, there are in fact very few states in the 
tail of a band of localized eigenfunctions 
\cite{sh-ef}. In Appendix C, 
we show that, for the simple lattice model of population growth 
discussed in Section 1,
\be
\Gamma_{gs} \approx a+ wd +\Delta 
\eqnum{2.9}
\label{eq:twonine}
\ee
and 
\be
g(\Delta \Gamma)\sim\exp[-({\rm const}./\Delta\Gamma)^{d/2}]
\eqnum{2.10}
\label{eq:twoten}
\ee
as $\Delta \Gamma\to 0$. Solving Eq. 
(\ref{eq:twoeight}) for large $R$ now leads to $\Delta \Gamma 
\sim 1/\ln(R/l_0)^{2/d}$ where $l_0$  is the lattice constant
of the model. 
The relaxation time for a system of size $R$ is then
\be
t^*(R) \sim \ln^{2/d}(R/l_0)\;, \qquad 
\qquad {\rm localized\  growth}\;.
\eqnum{2.11}
\label{eq:twoeleven}
\ee

The sparse population of growth rates near the ground state results 
in slow logarithmic growth of the relaxation 
time $t^*(R)$ with system size 
$R$. A very different size dependence 
results for the delocalized modes of the homogeneous model. 
For delocalized plane wave eigenfunctions 
described by a spectrum like 
(\ref{eq:fifteen}), there are many more states 
close to the ground state. When $\v=0$
the density of states $g(\Delta\Gamma)\sim(\Delta\Gamma)^{d/2-1}$
near the band edge in the delocalized limit, and the above argument 
leads to
\be 
t^*(R) \sim R^2\;,\qquad \qquad {\rm delocalized\  growth}.
\eqnum{2.12}
\label{eq:twotwelve}
\ee

\section{Interactions and a Delocalization Transition}

A complete analysis of the nonlinear ``interaction'' terms in Eq. 
(\ref{eq:two}), or its lattice equivalent Eq. (\ref{eq:eighteen}), 
is beyond the scope 
of this paper.  Some progress is possible, 
however, when only a few strongly 
localized growth eigenvalues near the band edge
of the Liouville operator have a positive real part.  
In this limit, it 
is easy to demonstrate that localized population 
dynamics with interactions differs 
considerably from the dynamics of the plane waves which describe 
population growth in a homogeneous environment.  We
shall also argue that the sharp mobility edge separating localized 
eigenfunctions from delocalized ones \cite{hat} 
implies a delocalization transition with 
increasing convective velocity 
or average growth rate in population dynamics.
This transition is clearly present in the linearized growth 
model, and we shall give arguments that it may be  present 
as well in steady-state population 
distributions described by Eqs. 
(\ref{eq:two}) and (\ref{eq:eighteen}).

Let us write Eq. (\ref{eq:two}) in the form
\be
{\partial c (\x,t)\over\partial t} = \cL c(\x,t)-b c^2(\x,t)
\eqnum{3.1}
\label{eq:threeone}
\ee
where $\cL$ is the Liouville operator (\ref{eq:thirteen}), 
and study the steady state which develops for long times.  
Upon expanding in the complete set of right eigenfunctions of 
$\cL$, with eigenvalues $\Gamma_n$
\be
c(\x,t) = \sum_n c_n(t)\phi^R_n(\x)\;,
\eqnum{3.2}
\label{eq:threetwo}
\ee
the dynamical equations read 
\be
{dc_n(t)\over dt}=\Gamma_n c_n(t)-\sum_{m,m'}
w_{n,m m'}c_m (t)c_{m'}(t),
\eqnum{3.3}
\label{eq:threethree}
\ee
where the mode coupling coefficients are
\be
w_{n,mm'}=b\int d^d \x\phi^L_n(\x)\phi^R_m(\x)\phi^R_{m'}(\x)
\eqnum{3.4}
\label{eq:threefour}
\ee
Upon combining the ``gauge transformation'' 
Eq. (\ref{eq:seventeen}) with 
the approximate form (\ref{eq:twofour}) 
of the localized eigenfunctions, the 
orthogonality condition (\ref{eq:twothree}) 
leads to \cite{hat}
\begin{eqnarray}
\phi^R_n& \simeq &\sqrt{{( 2 \kappa_n)^d \over \Gamma(d) \Omega(d)}}
\exp[\v \cdot (\x - \x_n)/D - \kappa_n |\x-\x_n|]  \nonumber \\    
\eqnum{3.5}
\label{eq:threefive}     \\
\phi^L_n& \simeq &\sqrt{{( 2 \kappa_n)^d \over \Gamma(d) \Omega(d)}}
\exp[-\v \cdot (\x - \x_n)/D - \kappa_n |\x-\x_n|]  \nonumber 
\end{eqnarray}
where  $\Omega_d$ is the surface area of a 
unit sphere in $d$ dimensions.

Now consider what happens in, say, 
the lattice population growth model 
(\ref{eq:eighteen}) when we vary 
the average growth rate $a$, starting with 
large negative values.  As discussed 
in Appendix C, the real part of the 
growth spectrum broadens to a width of order $wd 
+ \Delta $ about $a$ due to diffusive hopping between sites and 
randomness. If $-a \gg wd+\Delta$, then the 
real parts of all 
eigenvalues in Eq. (\ref{eq:threethree}) 
are negative and the population becomes extinct.  
As we increase $a$ however
we eventually reach the simple but interesting 
situation where a small fraction of the 
eigenvalues become positive.  
The coupled equations 
(\ref{eq:threethree}) then resemble the set of 
renormalization group recursion relations governing 
flows in the  space of 
Hamiltonian coupling constants which describe equilibrium 
critical points to one loop order  \cite{wilson}. Here, an expansion 
in $\epsilon = 4-d $ is used to truncate an infinite set of 
coupled differential equations.  By 
analogy with critical phenomena, we call
the modes with positive ${\rm Re} (\Gamma_n)$ ``relevant variables'' 
and those with negative  ${\rm Re} (\Gamma_n)$ 
``irrelevant variables''. 
To a first approximation, we can simply discard the irrelevant 
variables in Eq. (\ref{eq:threethree}), 
because their negative eigenvalues are 
not much affected by the dilute concentration of 
growing modes (see
below), and they will eventually die out.

The behavior of the relevant variables is also simple.  
Because these constitute only a small fraction of the 
total number of localized modes, they will be 
widely separated in space.  The overlap 
integral which defines the coupling 
coefficients in Eq. (\ref{eq:threefive})
will then be negligible unless $m=m'=n$ and 
the differential equations describing the 
localized eigenmodes decouple,
\be
{dc_n(t) \over dt} = \Gamma_n c_n(t)-w_n c^2_n (t)
\eqnum{3.6}
\label{eq:threesix}
\ee 
where
\be
w_n \equiv w_{n,nn}=b\int d^d\x \phi_n^L(\x) [\phi_n^R(\x)]^2
\eqnum{3.7}
\label{eq:threeseven}
\ee
If the convective velocity is small, 
the eigenvalues $\Gamma_n$ remain locked at 
their values for $\v \propto \g = 0$, 
and the neglected terms are smaller by a power of
$\exp[-\kappa l]$, where $l$ 
is a typical spacing between relevant eigenmodes.  

The evolution of $c
(\x,t)$ at long times (after the irrelevant variables have  
died off) is determined by substituting the solutions 
of Eq. (\ref{eq:threesix}),
\be
c_n(t) = {c_n(0) e^{\Gamma_n t} \over 1+c_n(0) 
{w_n \over \Gamma_n} 
(e^{\Gamma_n t} -1)},
\eqnum{3.8}
\label{eq:threeight}
\ee
into Eq. (\ref{eq:threetwo}) and only 
summing over the unstable modes.  
In contrast to the homogeneous population dynamics,  
where a {\it single} $\k = 0$ eigenfunction 
completely dominates the steady 
state (see Appendix B), the fastest growing localized 
eigenfunction does not interact appreciably with the other relevant 
variables.  The steady-state fixed point $c^*(\x)$ approached at long 
times is then characterized by {\it many} occupied modes,
\be
c^*(\x) = \mathop{{\sum_n}'} c_n^*\phi_n^R(\x) 
\eqnum{3.9}
\label{eq:threenine}
\ee
where the fixed point values $c_n^*$ are
\be
c^*_n = {\Gamma_n  \over w_n}\;,
\eqnum{3.10}
\label{eq:threeten}
\ee
and $\sum_n'$  means only  
unstable eigenfunctions make a nonzero contribution to the sum.  
This state is similar to the ``Bose glass'' 
phase of flux lines in Type 
II superconductors, where $a$ plays a role of a chemical potential 
for vortices \cite{nel-vin}. Unlike the Bose glass, 
however, the number of degrees of freedom associated with each 
occupied localized state is highly variable.  The ``occupation 
number'' $N_n$ of an unstable eigenmode is, from 
Eq. (\ref{eq:threenine}),
\be
N_n = c_n^* \int d^d\x \  \phi^R_n(\x)
\eqnum{3.11}
\label{eq:threeleven}
\ee

\begin{minipage}[t]{3.2in}
\epsfxsize=3.2in
\epsfbox{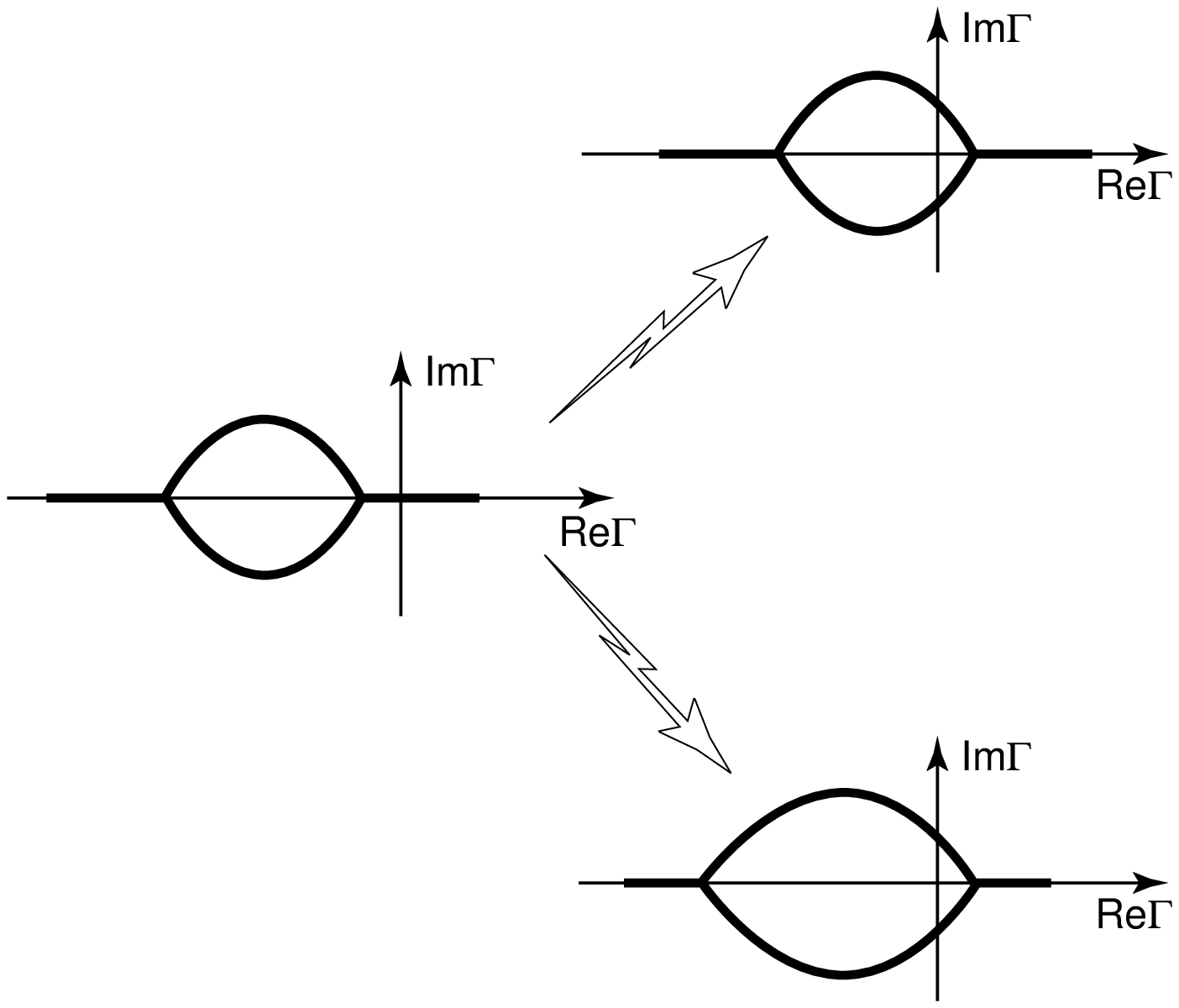}
\begin{small}
FIG.\ 5. 
Schematic of a one-dimensional eigenvalue spectrum 
below and above the delocalization threshold. On the left, 
only a few localized states are unstable and the bubble 
of ``irrelevant'' delocalized states has little 
effect on the steady state. Spectra on the right have 
unstable delocalized states arising either from an increase
in the average growth rate $a$ (top) or an increase in the 
convective velocity $v$ with $a-v^2/4D$ held fixed (bottom).
\end{small}
\vspace{0.15in}
\end{minipage}

We checked the above analysis in one dimension for $g\propto v=0$ by  
numerically determining the eigenfunctions in Eq. (\ref{eq:twenty}) 
for a particular realization of the random potential 
$U (x)\in[-\Delta,\Delta]$ with $a$ sufficiently 
negative so that only
four eigenvalues on a 700 site lattice were positive.  This  
situation is qualitatively similar to that 
shown on the left side of 
Fig. 5, except that there is no bubble of  
delocalized states in the center of the band for 
$g=0$. However, provided 
${\rm Re}(\Gamma_n)$ is large and negative for these delocalized
states, they rapidly die off and will not affect the fixed point 
describing the steady state.  We then determined the steady 
state population distribution for the full nonlinear equation 
(\ref{eq:eighteen}) under the same conditions.  This steady state 
is compared with the 
four relevant eigenmodes of the linearized problem in Fig. 6.  
The peaks in the exact steady state (top part of the 
figure) do indeed occur precisely at 
the locations of the four unstable growth eigenfunctions.  
It is easy to see from Eq. (\ref{eq:threefour}) 
that $w_{n,nn}={\rm const.}\; \kappa_n^{
d/2}$.  Since $\kappa_n$ (the inverse width of the localized 
eigenfunctions) is approximately independent of $n$ and equal 
to the lattice constant in the tail of the band (see, e.g.,
Ref. \cite{hat}), it follows 
from Eq. (\ref{eq:threeten})  
that the {\it heights} of the peaks in the steady state
 are proportional to the growth {\it eigenvalues} of the unstable 
modes in this simple model of population dynamics.  
We have checked that this 
relationship between peak heights and growth 
rates   
is satisfied by the steady state of     

\begin{minipage}[t]{3.2in}
\epsfxsize=3.2in
\epsfbox{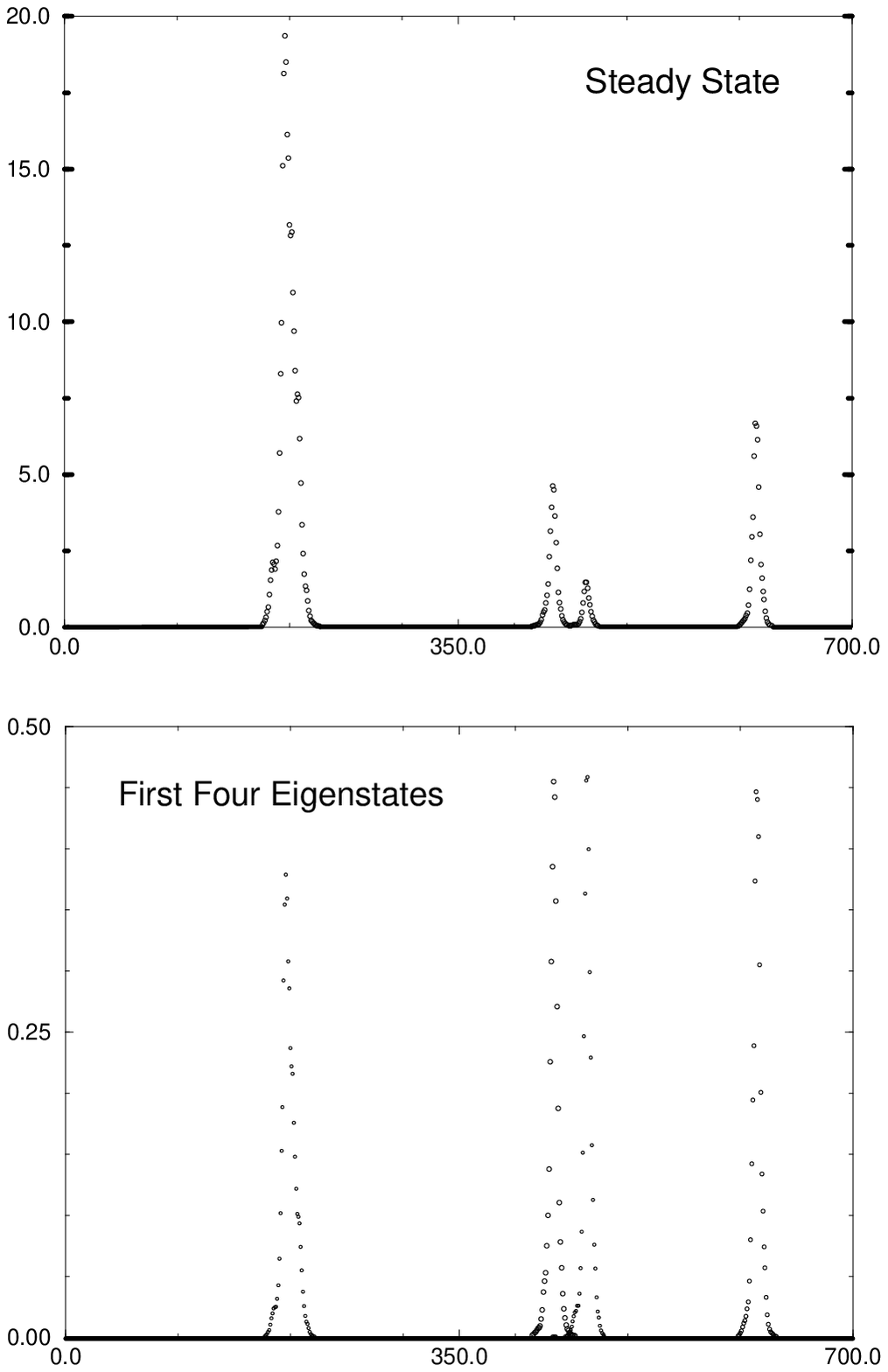}
\begin{small}
FIG.\ 6. 
(Top) Steady-state population distribution for a 
700-site tight-binding model with site random growth 
rates and nonlinear interactions. (Bottom) Plots of the 
four unstable eigenfunctions obtained by linearizing the 
Liouville operator for the same realization of the random 
potential about the state of zero population. The positions 
of these eigenfunctions match perfectly the peaks in the 
steady state of the nonlinear problem. The {\it heights}
of the peaks in the steady state are proportional to the 
{\it eigenvalues} of the corresponding eigenfunctions.
\end{small}
\vspace{0.25in}
\end{minipage}

\noindent
Fig. 6
with an accuracy of a few percent.  Thus, a 
population of bacteria described by 
(\ref{eq:two}) or (\ref{eq:eighteen}) 
evolves toward a steady-state 
distribution given by the ground state and first 
few excited states of a Schr\"odinger-like equation. 

We next discuss the case shown schematically 
on the right side of Fig. 5, where  
a number of {\it delocalized} states have become unstable.  
This regime could be accessed either by increasing the mean growth
rate $a$, or by increasing $\v$ which enlarges the 
bubble of extended states. (To pin the center of the 
band one should actually increase $v$ while holding $a-v^2/4D$ fixed). 
Because the Liouville operator is real, these delocalized 
modes occur in pairs: if $\phi_n^R(x)$ is a mode with 
eigenvalue $\Gamma_n$, then $\phi_n^R(x)^*$ is an eigenfunction
with eigenvalue $\Gamma_n^*$. As shown in detail by Brauwer
{\it et al.} \cite{brauwer}, these modes  
are (at least for weak randomness) approximately plane
waves, characterized by nonzero wave vector pairs $k$ and 
$-k$.  
The dynamics changes dramatically as soon as the first pair
of delocalized eigenfunctions becomes relevant. Now, there 
will be nontrivial mode couplings between the newly 
unstable delocalized modes and each other, as well as with 
the unstable 
localized ones discussed earlier. By analogy with the physics 
of tilted vortex lines interacting with columnar defects 
\cite{nel-vin,hwa}, we now expect macroscopic occupation of modes
near $q=0$, similar to Bose-Einstein condensation. A broad
background {\it extended} species population should now be 
superimposed on the peaks which represent localized modes,
as indicated schematically in Fig. 7. The proportion
of, say, bacteria incorporated into this background should
increase with increasing $a$. Delocalization arises because 
populations  can drift between growth ``hot spots'' instead of  
dying out when the  convection velocity is high.
According to the mode couplings in the Fourier basis 
displayed in Eq. (\ref{eq:bthree}) of Appendix B, fixed point value 
for the $q=0$ mode is determined by the values of the 
unstable pairs $(k,-k)$.

There should be  a large difference in the response of the 
steady state to a small change in $\v$ for localized 
and delocalized steady states. The spectrum ``unzips''
further into the complex plane with increasing $v\propto g$ as
more modes delocalize. For spectra like  that on the left
side of  Fig.~5, 
these additional delocalized modes are irrelevant and should not 
affect the steady state. The only change occurs due to the 
distortion of the relevant localized modes according to 
Eq. (\ref{eq:threefive}). For spectra like those on the 
right side of Fig.~5, 
however, increasing $\v$ (with $a-v^2/4D$ constant) 
leads to more {\it relevant} 
delocalized modes, with large changes in the corresponding 
steady state.
Note that the ``carrying capacity'' defined by Eq. 
(\ref{eq:threeleven}) {\it diverges} when $g\rightarrow
\kappa_n$ and a mode described by an eigenfunction like 
(\ref{eq:threesix}) becomes delocalized. However, 
coupling coefficients such as (\ref{eq:threeseven}) remain
{\it finite} at the delocalization transition.

Although we have discussed the dynamics using one-dimensional
spectra, we expect similar delocalization phenomena with 
increasing $v$ or $a$ with two-dimensional spectra such 
as that in Fig. 3b. If $a$ is adjusted so that only a few 
localized modes are relevant, and the growth modes are 
strongly localized, the steady state should look like 
Fig. 4 with peak heights proportional to the growth 
eigenvalues. With increasing growth rate or convection 
velocity, eventually both delocalized {\it and} new 
localized modes should start to participate in the steady 
state. The coexistence of localized and extended eigenmodes
at the same real part of the energy in $d=2$ was discussed
in \cite{hat}: anisotropic localized eigenfunctions for 
$\g=0$ will delocalize sooner if their most extended
direction coincides with the direction of $\g$. We expect 
that the resulting extended state wave functions are 
streaked out in the direction of $\g$, and that these 
streaks will be reflected into the steady-state species 
population.

\begin{minipage}[t]{3.2in}
\epsfxsize=3.2in
\epsfbox{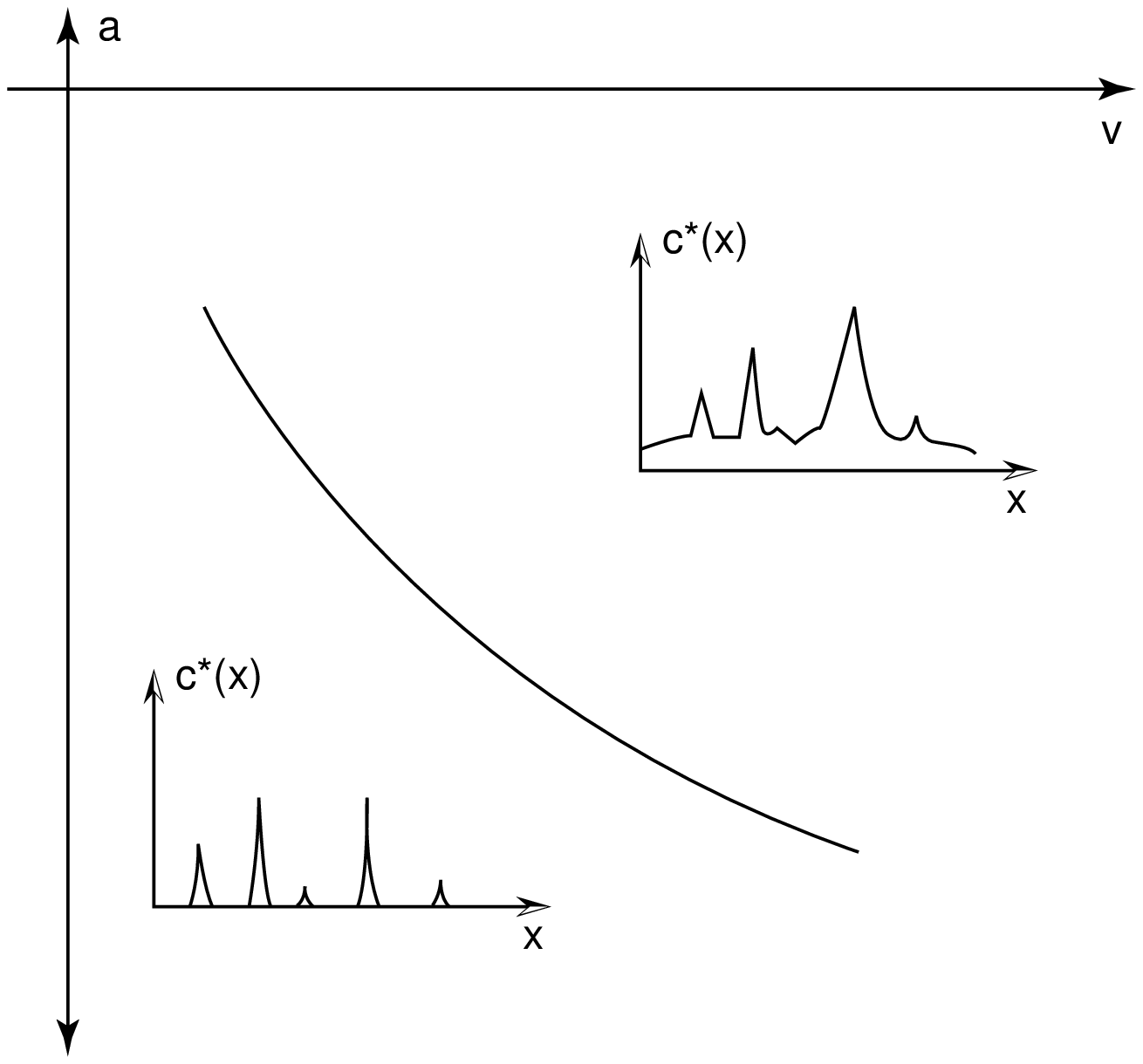}
\begin{small}
FIG.\ 7. 
Schematic ``phase diagram'' indicating 
regimes of localized steady states (lower left) and 
extended steady states (upper right) as a function of the 
average growth rate $a$ and the convection velocity $v$. (A
region of large negative $a$ where all initial 
populations eventually become extinct is not shown.)
\end{small}
\vspace{0.in}
\end{minipage}

\bigskip

It is possible that a delocalization transition arises 
even for $\v=0$ with increasing $a$ (or increasing 
diffusion constant $D$). As the number or spatial extent of 
the relevant 
eigenmodes goes up, eventually the localized states begin to 
overlap. The steady state may become ``extended'' in this 
limit, in the sense that individual members of a species 
can hop easily from one growth hot spot
to another. Again, the response 
of the steady state to a small change in convection velocity (away
from zero) could be an indicator of this transition. A sharp
transition is suggested by analogy with the ``Bose glass''
transition which occurs for increasing magnetic fields for 
vortices in Type II superconductors with columnar defects 
(see Appendix A  and Ref. \cite{nel-vin}).

A more detailed discussion of these interesting 
delocalization transitions will be presented in a future 
publication \cite{shnerb}. However, in the next two sections, 
we shall make some progress in describing the low-lying 
states which describe the  growth modes in 
$d=2$ for large convective velocities.

\section{Burgers' Equation and the Limit of Large $\v$}
\subsection{Qualitative Discussion}

We now study ``unbounded growth'' in the limit of large
$\v$, with the goal of better understanding complicated
spectra like those for the two-dimensional linearized 
growth problem shown in Fig. 3c. By large $\v$ we mean 
velocities so large that even the ``ground state,'' i.e.,
the growth mode with the largest eigenvalue, is 
delocalized,
\be
v>>D/\xi_0
\eqnum{4.1}
\label{eq:fone}
\ee
where $\xi_0$ is the localization length of the most 
rapidly growing eigenfunction. In this regime, it is helpful to 
exploit the analogy with the equilibrium statistical mechanics of a
$2+1$-dimensional vortex line described in Appendix A. In this analogy, 
a vortex trajectory represents the path taken by a particular 
growing population of, say, bacteria.  As is 
evident from Fig. 10, when $v\rightarrow\infty$ the 
component of the 
``magnetic field'' tilting the equivalent elastic vortex line leads
to configurations more nearly perpendicular to the columnar
defects representing the disordered growth rates.
The transverse fluctuations of the tilted vortex 
line's trajectory are unimpeded in the $\tau$ directions, while 
in the $xy$ plane it sees the cross sections of the columnar
defects. One might guess that in this large tilt limit the 
vortex simply wanders diffusively along  the 
$\tau$ direction, but acts like a directed polymer in a 
$1+1$-dimensional medium with point-like disorder when 
projected into the $xy$ plane. More generally, the physics of 
tilted vortex lines with columnar pins in $d+1$ dimensions 
should be related to directed polymers in $(d-1)+1$ 
dimensions with point-like disorder. A great deal is known
about these problems \cite{halpin}. 

To apply similar ideas to linearized population growth 
models, we assume that $\v$ is in the $x$ direction, and 
rewrite Eq. (\ref{eq:sixteen}) as
\be
\partial_t c+v\partial_xc=
D\partial_x^2c+D\nabla_\perp^2c+
[a+U(x,\r_\perp)]c
\eqnum{4.2}
\label{eq:ftwo}
\ee
where $\r_\perp$ represents all spatial coordinates perpendicular
to $x$. Specifically, we explore the long-time dynamics
generated by this equation with a delta function initial 
condition,
\be
\lim_{t\rightarrow 0}
c(x,\r_\perp,t)=
\delta(x)\delta^{d-1}(\r_\perp)
\eqnum{4.3}
\label{eq:fthree}
\ee
corresponding to a point inoculation of population at 
the origin. We look for a delocalized solution valid for 
long times in the limit of large $v$. The overall 
exponential time dependence generated by $a$, the constant 
part of the growth rate, and the diffusion with drift we
expect in the $(x,t)$ variables may be incorporated via
the substitution
\be
c(x,\rp,t)=
{e^{at}\over\sqrt{4\pi Dt}}
e^{{-(x-vt)^2\over 4Dt}}
W(x,\rp)
\eqnum{4.4}
\label{eq:ffour}
\ee
where $W(x,\rp)$ is to be determined. Note that $c(x,\rp,t)$
becomes proportional to $\delta(x)$ as $t\rightarrow 0$, so the 
initial condition (\ref{eq:fthree}) requires
\be
\lim_{x\rightarrow 0}
W(x,\rp)=\delta^{d-1}(\rp)\;.
\eqnum{4.5}
\label{eq:ffive}
\ee

Upon inserting Eq. (\ref{eq:ffour}) in (\ref{eq:ftwo}), we find

\begin{eqnarray}
v\partial_xW(x,\rp)&+&
{(x-vt)\over t}
W(x,\rp) \nonumber \\ 
&=&
D\partial_x^2W(x,\rp)+D\nabla_\perp^2
W(x,\rp)
\nonumber \\
&&+U(x,\rp)W(x,\rp)\;.
\eqnum{4.6}
\label{eq:fsix}
\end{eqnarray}
According to the ansatz (\ref{eq:ffour}), $c(x,\rp,t)$ is 
only appreciable for
\be
|x-vt|\; \lot \;2\sqrt{Dt}
\eqnum{4.7}
\label{eq:fseven}
\ee
so the second term on the left-hand side of (\ref{eq:fsix}) is 
smaller than the first by a factor of order $\sqrt{D/tv^2}$, and 
can be neglected in the limit of long times. In the remaining 
equation for $W(x,\rp)$, which has no explicit time 
dependence, we expect that the term $\partial_x^2 W(x,\rp)$ 
can be neglected for large $x$ and $t$ compared to the 
single $x$ derivative which appears on the left-hand side.
The resulting equation is an imaginary time Schr\"odinger 
equation, where $x$ plays the role of ``time'',
\be
v\partial_xW(x,\rp)\mathop{\approx}\limits_{t,x\rightarrow\infty}
D\nabla_\perp^2
W(x,\rp)+
U(x,\rp)W(x,\rp)\;.
\eqnum{4.8}
\label{eq:feight}
\ee
Note that the random ``potential'' $U(x,\rp)$ depends both 
on the ``time'' $x$ and on the additional $(d-1)$ spatial directions.

Consider the application of this mapping to 
two-dimensional species populations with strong convection.
Assume for simplicity that $a>0$, so that the population grows 
on average as it convects and diffuses downstream. For fixed $x$, the 
solution $W(x,y)$ of the resulting $(1+1)$-dimensional
Schr\"odinger equation (subject to the boundary condition
(\ref{eq:ffive}))
describes the distribution in $y$ of a growing species 
population which has traveled through random distribution 
of growth rates for a time of order $t = x/v$. The results of 
extensive studies of the  Schr\"odinger 
equation in $1+1$ dimensions \cite{halpin} with a 
space and time-dependent random potential 
may be interpreted as follows: For any fixed 
$y$ value, imagine tracing the genealogy of, say, all 
bacteria which have reached  particular 
position $(x,y)$. As $x \to \infty$, {\it the overwhelming majority 
of bacteria near the point $(x,y)$ will have 
evolved along a spatially convoluted  
optimal path of especially favorable growth rates. }
The fraction of bacteria whose ancestors come along this route is favored over
 all routes in $W(x,y)$ by a exponential factor $ \sim \exp[c'x^{\omega}]$, 
where $c'$ is a constant. The exponent $\omega$ (which describes the
fluctuations in the ground state energy in the analogous 
problem in the statistical mechanics of flux lines) in known 
to be  $\omega = 1/3$ exactly \cite{halpin}. Any particular 
path $y(x)$ of optimal evolution itself wanders with typical transverse 
fluctuations which behave like  
\be
y(x)\sim x^\zeta
\eqnum{4.9}
\label{eq:fnine}
\ee
where $\zeta= {\omega+1 \over 2} = 2/3$ is a universal critical exponent, 
independent
of the exact probability distribution of the randomness and other 
details. {\it Because $2/3 > 1/2$, this optimal path 
will be well defined even in the presence of diffusion.}
A schematic of a set of optimal paths is shown in Fig. 8a. The 
population distribution $c(x,y,t=x/v)$ is sketched in Fig. 8b. 
The exponent $\zeta$ also controls the overall transverse 
spread in  $y$ of the spatially varying population 
for fixed $x$. This ``superdiffusive'' spreading $(\zeta > 1/2)$ 
arises because trajectories which would be rare  in conventional diffusion 
lead to strong amplification if they pass through regions of particularly 
favored growth.   The exponent $\omega$ determines the {\it size} 
of the fluctuations in  $c(x,y,t=x/v)$, which ride on top 
of an overall exponential growth, $c(x,y,t=x/v) \sim e^{ax/v}$.   

Similar behavior is expected for the $2+1$-dimensional 
random Schr\"odinger equation which results for convecting populations
with randomness in $d=3$, with the universal exponents
$\zeta\approx 0.59\approx  3/5$ and $\omega = 2 \zeta -1 = 1/5$ \cite{halpin}.
 The growing population again  
becomes streaked-out in streamwise direction, but with a 
nontrivial wandering transverse to the stream.

\begin{minipage}[t]{3.2in}
\epsfxsize=3.2in
\epsfbox{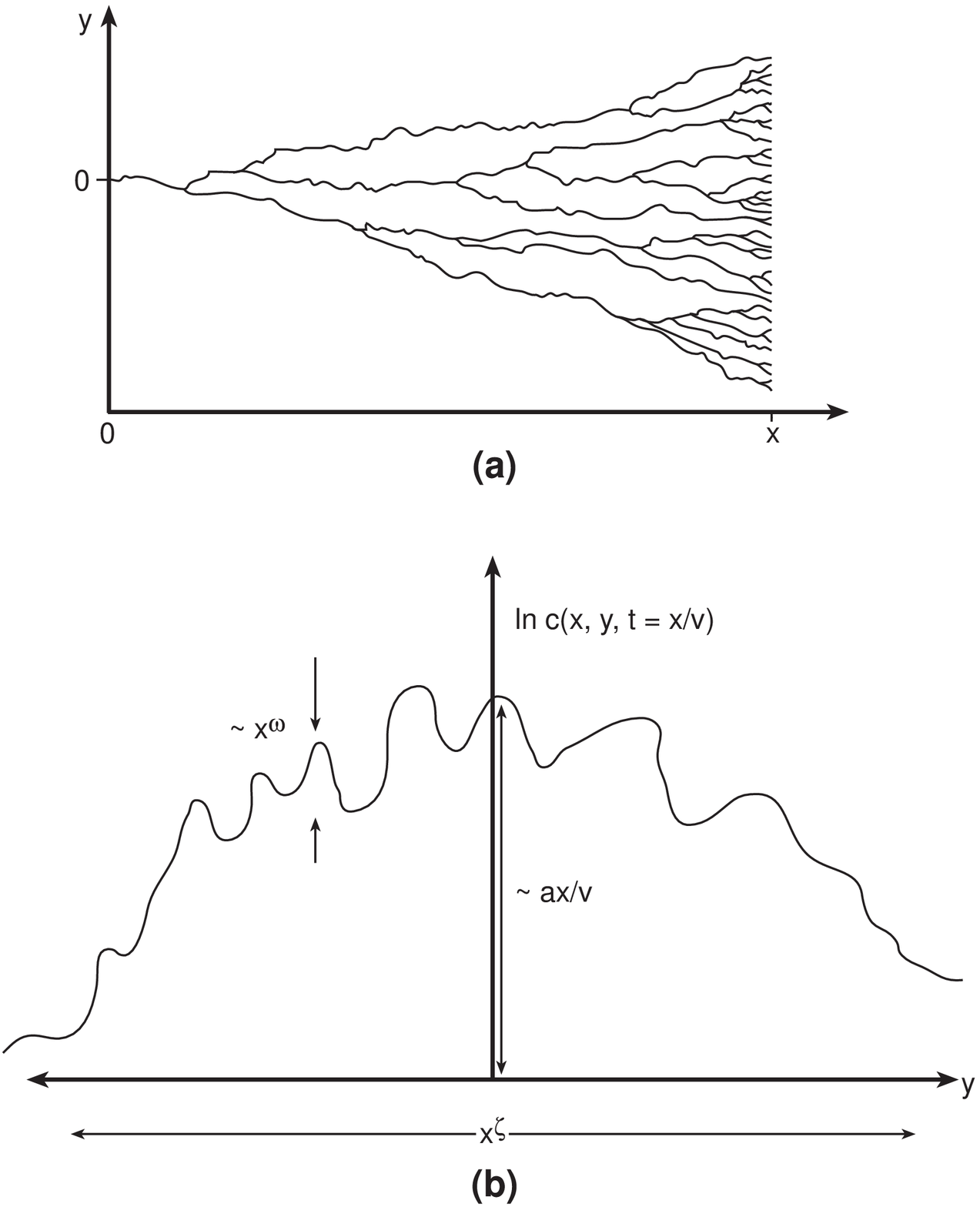}
\begin{small}
FIG.\ 8.
(a) Trajectories for a growing species 
which will produce a particularly large population at the point $(x,y)$ at 
time of order $t = x/v$. The population associated with a given 
point is spread over a region with typical size 
$\sqrt{Dx/v}$. For large $x$, bacteria which have traveled 
along such a path of favorable growth rates will dominate the population 
at $(x,y)$. (b) Schematic of the logarithm of the 
population $ c(x,y,t=x/v)$ discussed above as 
a function of $y$. Typical fluctuations in population size away from 
simple exponential growth are of order $\exp[c'x^{\omega}]$, and 
the population from an initial source of point inoculation 
has spread out a distance of order $x^\zeta$. 
\end{small}
\vspace{0.2in}
\end{minipage}

\subsection{Average Growth Spectrum}

We now study large $v$ growth spectra like those in Figs. 2c 
and 3c averaged over many realizations of the disorder. 
Reference \cite{hat} presents numerical evidence and 
qualitative arguments that the effect of disorder on the 
spectrum of the non-Hermitian operator (\ref{eq:thirteen})
in {\it one} dimension for large $v$ is in fact very 
weak \cite{brower}.
It was argued that the more complicated
chaotic eigenvalue spectra observed in $d=2$ were due to 
level repulsion of discrete eigenvalues in the complex
plane. Here we first 
show explicitly that perturbation theory in the disorder simply leads
to a ``free particle'' dispersion relation like Eq. 
(\ref{eq:fifteen}) with renormalized values of $a$, $v$, 
and $D$ in one dimension. We then demonstrate that
the same perturbation analysis 
is singular in higher dimensions,  consistent with 
the mapping onto a $(d-1)$-dimensional random Schr\"odinger
equation described above.

For a given configuration of growth rates, we apply standard 
second-order perturbation theory \cite{shankar} to the 
operator (\ref{eq:thirteen}). The resulting growth rate 
spectrum, starting with a plane wave set of basis functions
reads,
\be
\eps(\k)=\eps_0(\k)+
\langle\k|U|\k\rangle+
\sum_\q
{|\langle\q|U|\k\rangle|^2\over
\eps_0(\k)-\eps_0(\q)}
+O[U^3(\x)]
\eqnum{4.10}
\label{eq:ften}
\ee
where for a basis appropriate to the lattice growth model
(\ref{eq:twenty}) with $N$ sites, we have
\be
|\k\rangle=
{1\over \sqrt N}
\sum_\x e^{i\k\cdot\x}
|\x\rangle,
\eqnum{4.11}
\label{eq:feleven}
\ee
and
\be
\eps_0(\k)=a-i\v\cdot\k-Dk^2+O(k^3)\;,
\eqnum{4.12}
\label{eq:ftwelve}
\ee
with
\be
\langle\q|U|\k\rangle
={1\over N}
\sum_\x e^{i(\k-\q)\cdot\x}
U(\x)\;.
\eqnum{4.13}
\label{eq:fthirteen}
\ee
The averages over disorder are easily calculated by first 
using this lattice model, and then passing to the continuum limit.
With $U(\x)$ uniformly distributed in the interval 
$[-\Delta,\Delta]$, we have
\be
\overline{U(\x)U(\x')}=
{1\over 3}\Delta^2
\delta_{\x,\x'}
\eqnum{4.14}
\label{eq:ffourteen}
\ee
where the overbar represents a disorder average. It follows that
\be
\overline{\langle\k|U|\k\rangle}=0
\eqnum{4.15}
\label{eq:ffifteen}
\ee
and
\be
\overline{
|\langle\q|U|\k\rangle|^2}=
\Delta^2/3N\;.
\eqnum{4.16}
\label{eq:fsixteen}
\ee
With $v$ in the $x$ direction, we average over disorder and take the 
limit $N\rightarrow\infty$ in Eq. (\ref{eq:ften}), and find
\begin{eqnarray}
\bar\eps(k_x, k_\perp)&=&
\eps_0(k_x,k_\perp)\;+\; {1\over 3}
l_0^d\Delta^2\int{dq_x\over 2\pi}  \nonumber \\
&&\times\int{d^{d-1}q_\perp\over
(2\pi)^{d-1}}
{1\over iv(q_x-k_x)+D(q^2-k^2)}  \nonumber  \\
\eqnum{4.17}
\label{eq:fseventeen} 
\end{eqnarray}
where $l_0$ is the lattice constant of a  lattice and we have
kept only the small $k$ expansion displayed in Eq. (\ref{eq:ftwelve}).

Evaluation of the average spectrum is particularly simple in one 
dimension, because the transverse wave vectors $\k_\perp$ and
$\q_\perp$ are absent. Upon extending the integration limits 
on $q_x$ to $\pm\infty$, setting 
$q_x-k_x=p_x$ and symmetrizing in $p_x$, we have
\begin{eqnarray}
\bar\eps(k_x) &=&a-ivk_x-Dk_x^2
+{1\over 3}l_0\Delta^2  \nonumber \\
&&\times \int_{-\infty}^\infty
{dp_x\over 2\pi}
{D\over D^2p_x^2-(iv+2Dk_x)^2}
\nonumber \\
&=& a-ivk_x-Dk_x^2+
{l_0\Delta^2i\over
6(iv+2Dk_x)}
\;.
\eqnum{4.18}
\label{eq:feighteen}
\end{eqnarray}
Expanding the disorder correction in $k$ leads to 
renormalized values of $a$, $v$, and $D$
\be
a_R=a+l_0\Delta^2/6v
\eqnum{4.19a}
\label{eq:fnineteena}
\ee
\be
v_R=v-l_0\Delta^2D/3v^2
\eqnum{4.19b}
\label{eq:fnineteenb}
\ee
\be
D_R=D+2l_0\Delta^2D^2/3v^3
\eqnum{4.19c}
\label{eq:fnineteenc}
\ee
We see that disorder increases the mean growth rate, decreases 
the effective convective velocity, and increases the effective 
diffusion constant in  an expansion in the 
dimensionless ratio $l_0\Delta^2D/v^3$.

The new growth eigenfunctions are slightly perturbed plane waves
\cite{shankar}
\be
|k\rangle'=
|k\rangle+
\sum_{q\not= k}
{\langle q|U|k\rangle\over
\eps_0(k)-\eps_0(q)}
|q\rangle
\eqnum{4.20}
\label{eq:ftwenty}
\ee
and are clearly still delocalized. Equation (\ref{eq:fnineteenb})
suggests that localization sets in whenever
\be
{l_0\Delta^2D\over v^3}
\got 1,
\eqnum{4.21}
\label{eq:ftone}
\ee
i.e., localization occurs when the effective drift velocity is 
renormalized to zero. 
The physical basis for this criterion is as follows. In the 
absence of convection, the {\it maximum} growth 
rate associated with the operator (\ref{eq:thirteen}) 
(i.e., the ground state eigenfunction 
of the corresponding Hamiltonian)
will be given by minimizing the variational function
\be
W[c(x)]=
{\int\left[ {D\over 2}
\left( {dc(x)\over dx} \right)^2
-{1\over 2} U(x)c^2(x) \right]
dx \over \int c^2(x)\;dx}.  
\eqnum{4.22}
\label{eq:fttwo}
\ee
The random potential $U(x)$ in the continuum limit now has
correlator
\be
\overline{
 U (x) U (x')}=
{1\over 3}\Delta^2l_0\delta(x-x')
\eqnum{4.23}
\label{eq:ftthree}
\ee
where $l_0$ is a microscopic cutoff, of order the lattice 
constant in a discrete growth model like (\ref{eq:twenty}).
We assume an exponentially localized nodeless ``ground
state'' growth eigenfunction centered on the 
origin $c_0(x)\sim\exp(-\kappa|x|)$, 
and replace the random part of (\ref{eq:fttwo}) 
by its root-mean square value. Upon neglecting
dimensionless coefficients of order unity, we find
\be
W(\kappa)\sim D\kappa^2-\Delta(l_0\kappa
)^{1/2}
\eqnum{4.24}
\label{eq:ftfour}
\ee
which is minimized for $\kappa=\kappa_0$, with
\be
\kappa_0\approx (\Delta l_0^{1/2}/D)^{2/3}\;,
\eqnum{4.25}
\label{eq:ftfive}
\ee
and we again  neglect constants of order unity.
The ``gauge transformation'' (\ref{eq:seventeen}) allows 
this state to remain localized only if $v/D\;\lot\;
\kappa_0$, which is equivalent to the criterion
(\ref{eq:ftone}).

We now demonstrate the singularities which arise for 
large $\v||\hat\x$ in higher dimensions $d\geq 2$. Upon 
setting
\be
p_x=q_x-k_x,\qquad \p_\perp=\q_\perp
\eqnum{4.26}
\label{eq:ftsix}
\ee
and symmetrizing in $p_x$, Eq. (\ref{eq:fseventeen}) becomes
\begin{eqnarray}
\bar\eps&(&k_x,\k_\perp)=
a-ivk_x-D_xk_x^2-D_\perp k_\perp^2
\nonumber \\
&&+{1\over 3}\Delta^2
l_0^d\int{dp_x\over 2\pi}
\int{d^{d-1}p_\perp\over(2\pi)^{d-1}} \nonumber \\
&&\times 
{D_xp_x^2+D_\perp(p_\perp^2-k_\perp^2)\over
[D_xp_x^2+D_\perp(p_\perp^2-k_\perp^2)]^2
-p_x^2(iv+2D_xk_x)^2}
\eqnum{4.27}
\label{eq:ftseven}
\end{eqnarray}
where we have separated the diffusion term in $\eps_0
(k_x,k_\perp)$ into components parallel and perpendicular 
to $\v$ with $D_x=D_\perp=D$. 
It is tedious but straightforward to demonstrate
that $a$, $v$, and $D_x$ suffer only finite 
renormalizations similar to Eqs. (4.19) when this formula 
is expanded in $k_x$. However, upon setting $k_x=0$ and we find
a diverging renormalization of $D_\perp$ in the long 
wavelength part of the integral,
\be
D_\perp^R=D_\perp+
{\Delta^2l_0^d\over 12v}
\int{d^{d-1}p_\perp\over (2\pi)^{d-1}}
{1\over p_\perp^2}+\;\;
\hbox{less singular terms}.
\eqnum{4.28}
\label{eq:fteight}
\ee

Consider the meaning of this infrared divergence for $d=2$. 
In a finite system of spatial extent $L_y\ll L_x$ 
in the $y$ direction,
we have
\be
D_\perp^R\approx D_\perp
\left[1+\alpha
{\Delta^2l_0^2\over vD_\perp}
L_y\right]
\eqnum{4.29}
\label{eq:ftnine}
\ee
where $\alpha$ is a positive dimensionless coefficient.
Thus a ``free particle'' spectrum of the form
\be
\bar\eps_0(k_x,k_y)\approx
a_R-iv_Rk_x-D_x^R
k_x^2-D_\perp^Rk_y^2
\eqnum{4.30}
\label{eq:ftzero}
\ee
is only a good approximation
provided $v$ is large enough such that
\be
{\Delta^2l_0^2\over
vD_\perp}L_y\;\lot\;1\;.
\eqnum{4.31}
\label{eq:fttone}
\ee
Equation (\ref{eq:fttone}) is consistent with numerical results
and a criterion based on a level repulsion argument for the 
lattice model \cite{hat}. However, for any fixed value of $v$,
there will {\it always} 
be nontrivial changes in the growth spectrum for 
sufficiently large $L_y$, consistent with chaotic 
spectra like that exhibited in Fig. 3c. The mapping
onto the physics of a $(d-1)$-dimensional random 
Schr\"odinger equation suggests that, when evaluated to all 
orders in perturbation theory, the renormalized wave vector-dependent
transverse diffusion constant actually diverges as $q_\perp
\rightarrow 0$ \cite{halpin}, $D_\perp^R(q_\perp)\sim
q_\perp^{1/\zeta-2}$, so that the disorder averaged
renormalized spectrum takes the form, valid for small
wave vectors,
\be
\bar\eps(k_x,k_\perp)\approx a_R-iv_Rk_x-D_x^R
k_x^2-A_\perp k_\perp^{1/\zeta}\;.
\eqnum{4.32}
\label{eq:ftttwo}
\ee
With $\zeta=2/3$, we have 
\begin{eqnarray}
\bar\eps(k_x,k_y)\approx
a_R-iv_Rk_x&-&D_x^Rk_x^2-A_\perp
k_y^{3/2} \nonumber \\
&&\hbox{(two dimensions)}\;.
\eqnum{4.33}
\label{eq:fttthree}
\end{eqnarray}
This growth spectrum implies that length scales in the 
transverse direction scale with the 2/3 power of 
streamwise length scales, consistent with Eq. (\ref{eq:fnine}).
The exponent 3/2 controlling transverse fluctuations also 
appears in the  renormalization group treatment of 
fluctuations in $ln c(r,t)$ presented in Sec.~5. In $d=3$, the 
mapping onto a random Schr\"odinger equation
leads to an exponent $\zeta\approx 3/5$, and
the small wave vector form,
\begin{eqnarray}
\bar\eps(k_x,k_\perp)\approx a_R-iv_Rk_x&-&D_x^R
k_x^2-A_\perp k_\perp^{5/3}\nonumber \\
&&{\rm (three\ dimensions)}.
\eqnum{4.34}
\label{eq:fttfour}
\end{eqnarray}

It is interesting to compute the consequences of spectra 
like (\ref{eq:fttthree}) and (\ref{eq:fttfour}) for a 
density of states in the complex plane, defined by
\begin{eqnarray}
g(\eps_1,\eps_2)&=&\int
{dk_x\over 2\pi}
\int{d^{d-1}k_\perp\over (2\pi)^{d-1}}
\delta[\eps_1\nonumber \\
&&-{\rm Re}\;\bar\eps(k_x,k_\perp)]
\delta[\eps_2-{\rm Im}\;\bar\eps(k_x,k_\perp)].
\eqnum{4.35}
\label{eq:fttfive}
\end{eqnarray}
Note that this is the density of states associated with 
disordered averaged {\it  eigenvalues}. If the fluctuations 
of the eigenvalues away from their average values are small, this 
quantity will be 
the same as the disordered 
average {\it density of states} studied, e.g., in Ref. \cite{janik}. 
We measure energies relative to $a_R$ so that the 
top of the band (corresponding to the ground state of the 
equivalent Hamiltonian) occurs near the origin. Straightward
calculations then lead to the predictions
\begin{eqnarray}
g(\eps_1,\eps_2)&=& 0,\quad
\eps_1>0\quad {\rm or}\quad \eps_2^2>
-\eps_1v_R^2/D_x;
\nonumber \\
g(\eps_1,\eps_2)&\propto&(|\eps_1|-D_x^R
\eps_2^2/v_R^2)^{-1/3}, \nonumber \\
&&\quad{\rm otherwise\ (two\ dimensions)}
\eqnum{4.36}
\label{eq:fttsix}
\end{eqnarray}
and
\begin{eqnarray}
g(\eps_1,\eps_2)&=& 0,\quad
\eps_1>0\quad {\rm or}\quad \eps_2^2>
-\eps_1v_R^2/D_x;
\nonumber \\
g(\eps_1,\eps_2)&\propto&(|\eps_1|-D_x^R
\eps_2^2/v^2)^{1/5}, \nonumber \\
&&\quad{\rm otherwise\ (three\ dimensions)}
\eqnum{4.37}
\label{eq:fttseven}
\end{eqnarray}
close to the band edge.

These results are valid for small real and imaginary 
energies $\eps_1$ and $\eps_2$. The density of states 
thus vanishes outside a parabolic boundary, as in
Fig.~3c and similar to its behavior in a pure system. 
However, it diverges as the boundary is approached in 
$d=2$ and {\it vanishes} in $d=3$ with new universal 
critical exponents when disorder is present. Disorder 
thus has a strong influence in the thermodynamic limit 
for large $v$ when $d\geq 2$, even though all states 
are delocalized. In a {\it homogeneous} growth model,
this density of states has a square root divergence 
near the boundary in $d=2$ and approaches a constant in 
$d=3$.

As a crude approximation,  one could use the {\it average}
growth spectrum to estimate the behavior of, for 
example $\bar c(x,y,t)$ in $d=2$,
\begin{eqnarray}
&\bar c&(x,y,t) \approx \nonumber \\
&& \int {dk_x\over 2\pi}
\int{dk_y\over 2\pi}
e^{\bar\eps(k_x,k_y)t}
e^{ik_xx+ik_yy}
c(k_x,k_y,t=0) 
   \eqnum{4.38}
   \label{eq:ftteight}
\end{eqnarray}
where $c(k_x,k_y,t=0)$ is the Fourier transformed initial 
condition. However, it is easier and more systematic to study the 
statistics of $\ln[c(x,t)]$, as is often case for systems 
with multiplicative noise. This is done in the next Section.

\section{Space-time Fluctuations of $\wnc$}

In this section we give a detailed analysis of the   response 
of a  homogeneous biological system to the introduction of 
quenched random inhomogeneities in the growth rate. Let us look, for 
concreteness, at a {\it homogeneous} initial condition, 
i.e.,  $c(\x ,=0) = {\rm const.}$, which then evolves under the 
influence of some kind of quenched spatial randomness. As
discussed in Section I,  if the initial, 
constant density  $c(\x,t=0)$
is very small compared to $a/b$, the short 
time dynamics of this system 
is determined by Eq. (\ref{eq:six}), and for $c(\x,t) \approx c^*(\x)$, 
i.e., near the stable fixed point, the long-time decay into $c^*$ 
is given by Eq. (\ref{eq:eight}). The dynamic 
renormalization group approach 
presented here is thus directly relevant to the 
decay of small unstable fluctuations into the steady state
or to the growth of unstable modes if $b$ is small enough
such that the long-time behavior of the 
linearized problem manifests itself 
before the nonlinear reaction term in (\ref{eq:two}) 
becomes important. 
Moreover, as discussed in Section III, the basic features of the 
linearized problem, such as its eigenvalues and eigenstates, 
can be used in certain limits 
to assess  the nonlinear time evolution, as 
in Eq.  (\ref{eq:threethree}).

We assume the term $-b c^2$ in 
(\ref{eq:two}) is negligible, and  
use the fact that $c(\x , t)$ is always nonnegative to define 
its logarithm  $\Phi(\x,t)$, via  the transformation
\be
c(\x,t) = e^{{\lambda \over 2 D} \Phi(\x,t) + at} c(\x,t=0)\;.
\eqnum{5.1}
\label{eq:fiveone}
\ee
With $\v\Vert\hat \x$, the function $\Phi(\x,t)$ then satisfies
(for a {\it uniform} initial species population)
\be
\partial_t \Phi + v \partial_x 
\Phi=D\nabla^2\Phi+{\lambda\over 2}
(
\nabla\Phi)^2 + U(\x)\;,
\eqnum{5.2}
\label{eq:fivetwo}
\ee
with the initial 
condition  $\Phi(\x,t=0) = 0$. 
An identical equation was studied numerically and via a 
scaling ansatz by Chen {\it et al.}  \cite{Chen}
as a model of charge density waves interacting with 
quenched disorder. Here, we use the 
renormalization group to study this 
problem analytically using the methods of Ref. \cite{Forster}. 

We have singled out the direction parallel 
to the drift as $x$, while the $d-1$-dimensional perpendicular 
space will be denoted  $\rp$, as in Sec.~4. 
Equation (\ref{eq:fivetwo}) takes the form 
\begin{eqnarray}
\partial_t \Phi(\rp,x,t) &+& v \partial_x 
\Phi(\rp,x,t) = \nonumber \\ 
& D_\perp & \nabla^2_\perp
\Phi (\rp,x,t) + D_x \partial^2_x
\Phi (\rp,x,t)  
\eqnum{5.3}
\label{eq:fivethree}  \\   
&+& \lambda_\perp ({\bf 
\nabla_\perp} \Phi)^2 + \lambda_x (\partial_x \Phi)^2 +  U(\rp,x)\;,
\nonumber
\end{eqnarray}
with $D_\perp=D_x=D$, and $\lambda_\perp=\lambda_x=\lambda$.
The random function $U(\rp, x)$ satisfies
\be
\overline{U(\rp, x)U(\r'_\perp,x)}=
\Upsilon \delta(x-x')\delta^{d-1}[\rp-\r'_\perp]
\eqnum{5.4}
\label{eq:fivefour}
\ee
where the correlator strength is related to the spread
$[-\Delta, \Delta]$ of growth rates in a lattice model
by $\Upsilon\propto\Delta^2\ell_0^d$.

We now impose   a change of scale
\begin{eqnarray}
\rp &\to&  s \rp 
\eqnum{5.5a}
\label{eq:fivefivea} \\ 
x &\to& s^\eta x  
\eqnum{5.5b}
\label{eq:fivefiveb} \\
t &\to& s^z t 
\eqnum{5.5c}
\label{eq:fivefivec} \\ 
\Phi &\to& s^\alpha \Phi 
\eqnum{5.5d}
\label{eq:fivefived}
\end{eqnarray}
where $s$ is a renormalization group scale factor.

Under this scale transformation the parameters of 
Eq. (\ref{eq:fivethree})
change according to
\begin{eqnarray}
D_\perp &\to& s^{z-2} D_\perp   
\nonumber \\   
D_x &\to& s^{z - 2 \eta} D_x 
\nonumber \\   
\Upsilon &\to& s^{2z-2\alpha -(d -1)-\eta} \Upsilon   
\eqnum{5.6}
\label{eq:fivesix}
\\
\lambda_\perp &\to& s^{z+\alpha-2} \lambda_\perp  
\nonumber \\ 
\lambda_x &\to& s^{z+\alpha- 2 \eta} \lambda_\perp 
\nonumber  \\
v &\to& s^{z-\eta} v 
\nonumber
\end{eqnarray}

If the nonlinearities are absent, i.e., 
$\lambda_x = \lambda_\perp = 0$, Eq. (\ref{eq:fivethree})
becomes exactly solvable.  In this case,  
$\Phi(\rp,x,t)$ is given by
\begin{eqnarray}
\Phi(\rp,x,t)&=& \int_{-\infty}^
{\infty} {d \omega \over 2 \pi}
\int_{-\infty}^{\infty}  {d k_x \over 2 \pi}   
\eqnum{5.7}
\label{eq:fiveseven}   \\
&\times & \int 
{d^{d-1} k_\perp  \over (2 \pi)^{d-1}} 
\Phi(k_\perp,k_x,\omega) e^{i \k_\perp \cdot \rp}  
e^{i k_x x} e^{-i \omega t} \nonumber
\end{eqnarray}
where the Fourier transform of $\Phi(\rp,x,t)$,  
satisfies
\be
\Phi(\k_\perp,k_x,\omega) =G_0(\k_\perp,k_x,\omega) U
(\k_\perp,k_x,\omega)
\eqnum{5.8}
\label{eq:fiveeight}
\ee
Note that since the disorder is time independent,
its Fourier transform satisfies  
\begin{eqnarray}
U(\q_\perp,q_x,\omega) = \tu(\q,q_x) 2 \pi  \delta(\omega), \nonumber
\end{eqnarray}
and
\be
\langle\tu(\k,k_x) \tu(\k',k_x')\rangle
= \Upsilon (2 \pi)^d \delta(\k_\perp-\k_\perp') 
\delta(k_x - k_x').  
\eqnum{5.9}
\label{eq:fivenine}
\ee
The bare propagator $G_0(\k_\perp,k_x,\omega)$ is 
\be
G_0(\k_\perp,k_x,\omega) = {1 \over -i 
\omega +ivk_x+ D_xk_x^2+ D_\perp k_\perp^2 }.
\eqnum{5.10}
\label{eq:fiveten}
\ee
In this case the naive scaling treatment gives exact results. 
>From Eqs. (\ref{eq:fivesix})
one finds that keeping $D_\perp$, $\Delta$ and $v$ fixed
under the scale transformation requires
\be
z = 2, \qquad  \eta = 2,  
\qquad \alpha = {3 - d \over 2}.
\eqnum{5.11}
\label{eq:fiveeleven}
\ee
The term proportional to $D_x$ then scales to zero, confirming
that it is negligible at long wavelengths and low frequencies
compared to the single derivative in the 
convective term. The same analysis suggests that the nonlinear
coupling $\lambda_x$ is also irrelevant, $\lambda_x\rightarrow
s^{-(1+d)/2}\lambda_x$, while $\lambda_\perp \rightarrow
s^{3-d/2}\lambda_\perp$ so that $d=3$ is the upper 
critical dimension for this problem. 

These results become easier to understand if we consider their 
impact on a concrete physical quantity. Following Ref. 
\cite{Chen}, we consider the sample-to-sample fluctuations 
in $\Phi(\r_\perp,x,t)$ in different random environments all with 
the same physical dimensions $L_x$ and $L_\perp$,
\begin{eqnarray}
W(L_x,L_\perp)&=& \overline{
\Phi^2(x,\rp,t)}
\nonumber \\
&=&\overline{
\ln^2 [c(x,\rp,t)e^{-at}]}\;.
\eqnum{5.12a}
\label{eq:fivetwelvea}
\end{eqnarray}
For population dynamics of, say, bacteria, one could divide 
a single large colony into many patches with these 
dimensions to calculate the average. It is straightforward 
to show that the simple scaling of the linear theory 
sketched above implies that $W(L_x,L_\perp)$ takes the 
form, for $d=2$
\be
W(L_x,L_\perp)=L_x^\chi h(L_\perp/L_x^\zeta)
\eqnum{5.12b}
\label{eq:fivetwelveb}
\ee
where $\chi=1/2$ and $\zeta=1/2$. To obtain (\ref{eq:fivetwelveb}), 
use Eq. (\ref{eq:fivetthree}) below with the Gaussian 
exponents from (\ref{eq:fiveeleven}). 
The same results follow 
directly from Eq. (\ref{eq:fiveeight}) \cite{Chen}
\be
W(L_x,L_\perp)=
\int_{q_x<L_x^{-1}}
{dq_x\over 2\pi}
\int_{q_\perp<L_\perp^{-1}}
{dq_\perp\over(2\pi)}
{\Upsilon\over
v^2q_x^2+D_\perp^2q_\perp^4}
\eqnum{5.12c}
\label{eq:fivetwelvec}
\ee
which also leads to the conclusion that $h(x)\sim x$ for large
$x$. Chen {\it et al.} simulated the full nonlinear equation
(5.2) in $d=2$ and found $\chi=0.5\pm 0.05$ and $\zeta=0.85\pm 0.05$. 
Below we determine these exponents exactly.

To consider systematically the effect of the nonlinearities,
we perform a perturbative expansion around the exact solution 
embodied in (\ref{eq:fiveseven}--\ref{eq:fiveten}) in powers 
of $\lambda_\perp$ and $\lambda_x$. We begin by 
rewriting (\ref{eq:fiveseven}) as an integral 
over the momentum  up to some cutoff:
\begin{eqnarray}
\Phi(\rp,x,t)& =& \int_{-\infty}^{\infty} 
{d \omega \over 2 \pi}
\int_{-\infty}^{\infty}  
{d k_x \over 2 \pi} \int_0^\Lambda 
{d^{d-1} k_\perp  \over (2 \pi)^{d-1}} \nonumber \\
&&\times 
\Phi(\k_\perp,k_x,\omega) c^{i \k_\perp \cdot \r_1}  
e^{i k_x x} e^{-i \omega t}
\eqnum{5.13}
\label{eq:fivethirteen}
\end{eqnarray}

\begin{minipage}[t]{3.1in}
\epsfxsize=3.in
\begin{small}
(a)
\end{small}

\epsfxsize=3.1in
\epsfbox{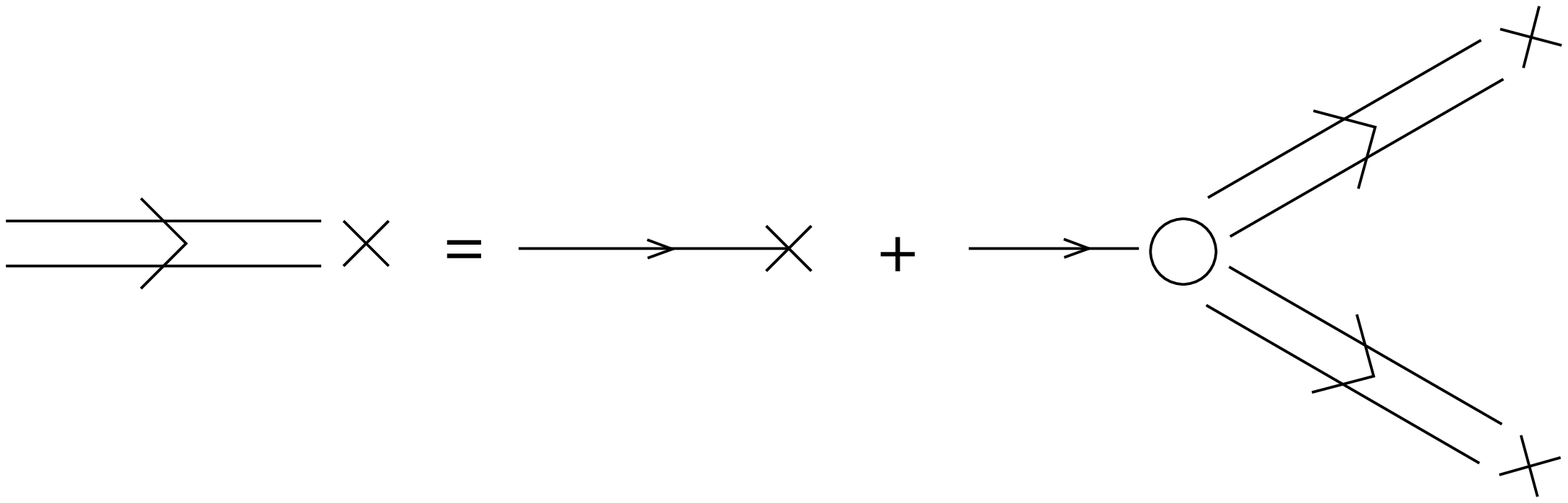}
\end{minipage}

\begin{minipage}[t]{3.1in}
\epsfxsize=3.1in
\begin{small}
(b)
\end{small}

\epsfxsize=3.1in
\epsfbox{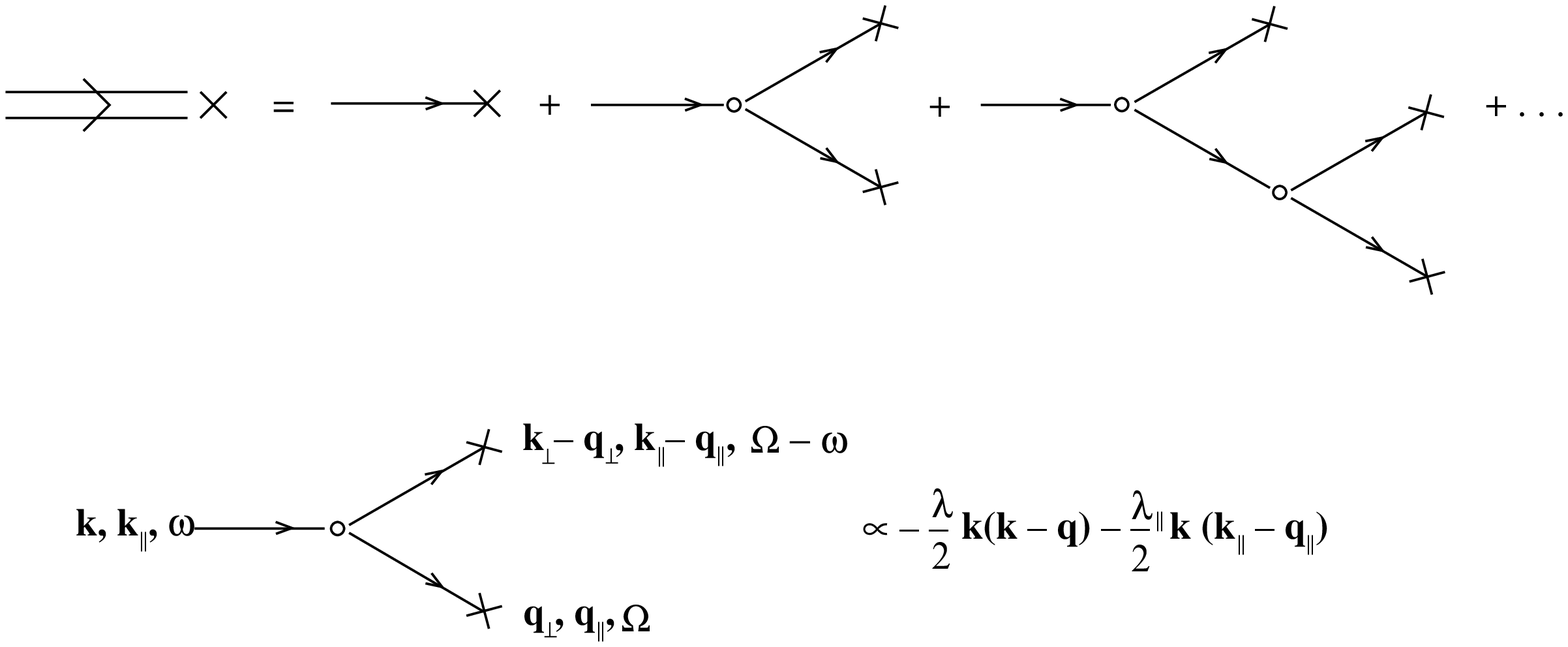}
\begin{small}
FIG.\ 9. 
(a) Diagrammatic representation of the integral equation
(\ref{eq:fivefourteen}). (b) Iteration solution of 
Eq. (5.14) and this meaning of the vertex in this 
diagrammatic series.
\end{small}
\end{minipage}

\begin{minipage}[t]{3.2in}
\epsfxsize=3.2in
\epsfbox{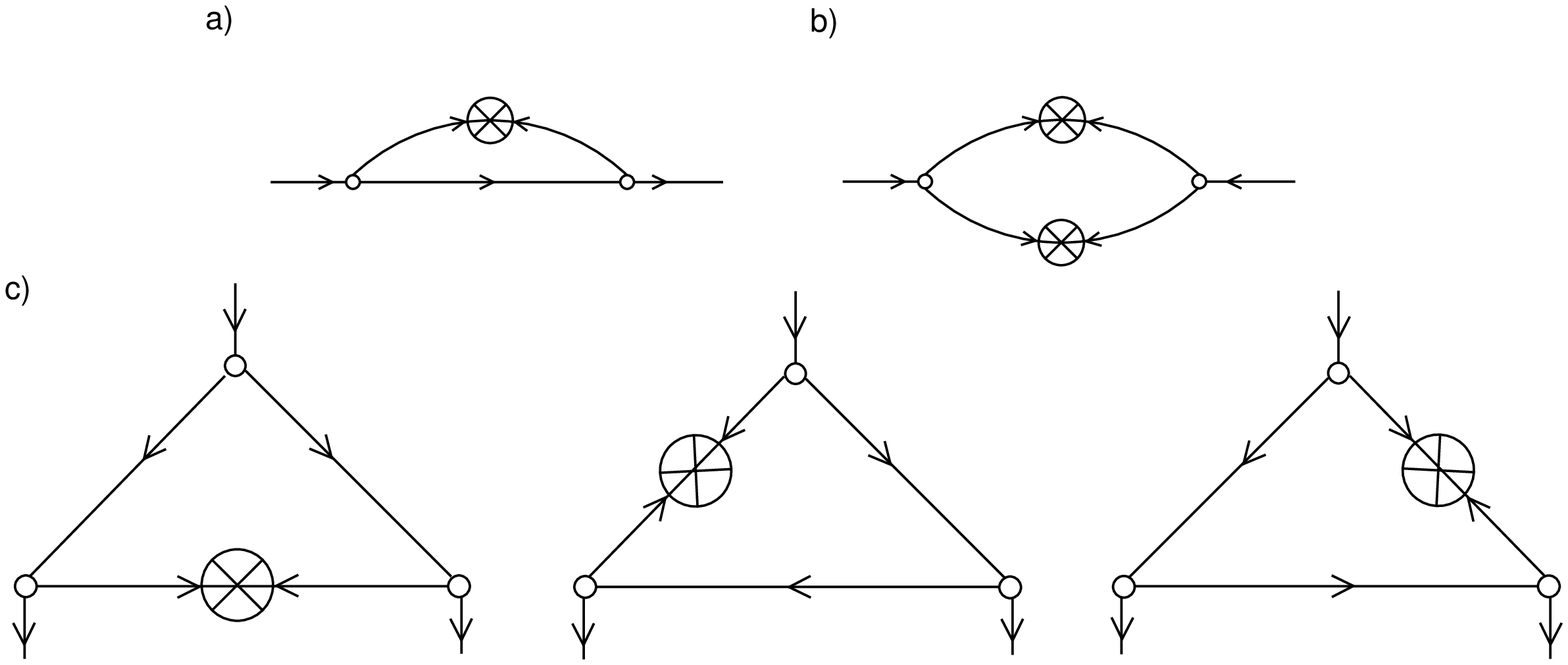}
\begin{small}
FIG.\ 10.
 (a) One-loop correction to the self-energy 
obtained by averaging over the noise. (b) Perturbative 
correction to the two-point function $\overline{\Phi
(\x,t)\Phi({\bf 0},0)}$.
(c) Three diagrams which contribute to the one-loop 
effective vertex  renormalization.
\end{small}
\vspace{0.25in}
\end{minipage}

\noindent
Here, $\Lambda$ is a cutoff initially of order $\ell_0^{-1}$,  
where $\ell_0$ is a microscopic length 
scale of order the lattice constant. 
Equation (\ref{eq:fivethree}) now becomes
\begin{eqnarray}
\Phi&&(\k_\perp,k_x,\omega)=G_0(\k_\perp,k_x,\omega) 
\left[ U(\q_\perp,q_x,\omega)   
+  \int_{-\infty}^{\infty} 
{d \Omega \over 2 \pi} \right. \nonumber \\
&&\int_{-\infty}^{\infty}  
{d q_x \over 2 \pi} \int_0^\Lambda 
{d^{d-1} \q_\perp  \over (2 \pi)^{d-1}} 
\nonumber   \\ 
&& \left( - {\lambda_\perp \over 2} \q_\perp \cdot
[\k_\perp-\q]_\perp
- {\lambda_x \over 2} q_x [k_x-q_x] \right) \nonumber \\     
&&\left.\Phi(\k_\perp-\q_\perp,k_x-q_x,\omega-\Omega)
\Phi(\q,q_x,\Omega) \right]
\eqnum{5.14}
\label{eq:fivefourteen}
\end{eqnarray}
which can be represented graphically as in Fig. 9a. 
The formal iterative solution 
of (\ref{eq:fivefourteen}) is shown in Fig. 9b.

The self-energy ${\BSigma}(\k_\perp,k_x,\omega)$ (Fig. 10a)
yields   the 
renormalized velocity $v_R$ and diffusion constants $D_\perp^R$ 
and $D_x^R$.   
The renormalization of  $\Upsilon$ is represented by the 
graph in Fig. 10b,  and vertex renormalization graphs 
for $\lambda_x$ and $\lambda_\perp$ are 
shown in Fig. 10c. The corrections 
to $\lambda_\perp$, $\lambda_x$, $D_x$ 
and $v$ do not diverge in the  infrared  
limit in any dimension;  naive perturbation
theory yields, however,  infrared divergent  corrections to  
$D$ and $\Upsilon$  for $d \leq 3$ as one takes the limit $\k \to 0$,
\be
\delta D = K_{d-1} 
\left({\lambda^2 \Upsilon \over 2 D^2 |v|}\right)
\left( {3-d \over 4d - 4} \right) 
\int_0^\Lambda dq q^{d-4}   
\eqnum{5.15}
\label{eq:fivefifteen}
\ee
\be
\delta \Upsilon = K_{d-1}  
\left({\lambda^2 \Upsilon \over 8 D^2 |v|}\right)
\int_0^\Lambda dq q^{d-4} 
\eqnum{5.16}
\label{eq:fivesixteen}
\ee
where $K_d = S_d/(2 \pi)^d$ and $S_d$ is the surface area 
of the $d$-dimensional unit sphere. 
Equation (\ref{eq:fivefifteen}) is similar to the result 
(\ref{eq:fteight}) obtained by considering the average
growth rate. 
Equations (\ref{eq:fivefifteen}) and (\ref{eq:fivesixteen}) 
confirm that 
the upper critical dimension of theory is $d=3$, below which 
renormalization group
techniques are needed to take care for the infrared divergences 
of the loop integrals. 
Upon carrying out the  
procedure of Ref. \cite{Forster}, we find 
that the renormalization group 
flow equations for the relevant variables  take the form
\begin{eqnarray}
{dD_\perp \over dl} &=& 
D_\perp\left[z-2+K_{d-1}\left({3-d\over4d-4}\right)g^2\right] 
\eqnum{5.17}
\label{eq:fiveseventeen} \\
{d\Delta \over dl} &=& \Delta  \left[2z - d +1   - 2 \alpha - \eta 
+ K_{d-1} {g^2 \over 4} 
\right] 
\eqnum{5.18}
\label{eq:fiveeighteen} \\
{d\lambda_\perp \over dl} &=& \lambda_\perp  
\left[\alpha + z - 2 \right]  
\eqnum{5.19}
\label{eq:fivenineteen} \\ 
{dv \over dl} &=& v  \left[ z - \eta  \right]  
\eqnum{5.20}
\label{eq:fivetwenty} 
\end{eqnarray}
We have set $s\equiv e^{-\ell}$, reduced the cutoff from $\Lambda$
to $e^{-\ell}\Lambda  $, and the corrections 
to the naive scaling results (\ref{eq:fivesix}) 
are proportional 
to the dimensionless 
coupling constant, 
\be
g^2 \equiv {\Delta \lambda^2 \over  2  D^3 |v|}\;.
\eqnum{5.21}
\label{eq:fivetone}
\ee 
The couplings $D_x$ and $\lambda_x$ again scale to zero, even at the 
nontrivial fixed point discussed below.

We now set $\eta=z$ and $\alpha=2-z$ to ensure that 
$\lambda_\perp$ and $v$ remain unchanged by our renormalization
procedure.
Using Eqs.  (\ref{eq:fiveseventeen})--(\ref{eq:fivetwenty})  
we can calculate the flow of the coupling constant $g$,
\be
{dg \over dl} = {3 - d \over 2}g +  
K_d \left({2d-5 \over  4d -4}\right) 
g^3. 
\eqnum{5.22}
\label{eq:fivettwo}
\ee
The fixed point $g^*$ is obtained by taking 
${dg \over dl} = 0 $. For 
$d = 2$ (i.e., two-dimensional disordered growth model with one 
parallel and one perpendicular 
direction) the Gaussian  fixed point at $g^*=0$ is unstable, while 
$g^* =  \left( {2 \over K_{d-1}} \right)$ 
is an attractive fixed point
which corresponds to $\alpha=1/2$, $z = 3/2$, 
and $\eta = 3/2$ 
($\eta$ should be equal to $z$ since there 
are  no infrared diverging 
correction to $v$). 

We now integrate the recursion relations 
until $\ell$ is large, so that 
all transients have died off. The renormalization group 
homogeneity relation for the mean square fluctuations in 
$ln [x(z,\rp, t)e^{-at}]$ in a system with dimension $L_x$
and $L_\perp$ then reads
\be
W(L_x,L_\perp)=e^{4\ell-2z\ell}
W(L_xe^{-z\ell},L_\perp e^{-\ell})
\eqnum{5.23}
\label{eq:fivetthree}
\ee
The prefactor follows from Eq. (\ref{eq:fivefived}),
with $\eta=z$ and $\alpha=2-z$; we expect these 
results to be correct to all 
orders in perturbation theory due to the Galilean invariance
of the problem after averaging over disorder \cite{halpin}.
Upon choosing $\ell=\ell^*$ such that $L_xe^{-z\ell^*}=1$, 
Eq. (\ref{eq:fivetthree}) can be rewritten in a scaling 
form similar to (\ref{eq:fivetwelveb}),
\be
W(L_x,L_y)=
L_x^{4\zeta-2} h(L_\perp/L_x^\zeta)
\eqnum{5.24}
\label{eq:fivetfour}
\ee
where
\be
\zeta=1/z\;.
\eqnum{5.25}
\label{eq:fivetfive}
\ee
We have thus confirmed the scaling ansatz Eq. (\ref{eq:fivefiveb})
\cite{Chen} with the specific predictions $\eta=1/z=2/3$ and 
$\chi=4\zeta-2=2/3$. These exact exponents differ somewhat,
however, from 
the numerical estimates of Chen {\it et al.} \cite{Chen}; 
it would be interesting to see if the agreement improves
with larger system sizes.

For $d=3$, there is no perturbatively accessible fixed point
of the one loop recursion relation (\ref{eq:fivettwo}). 
However, extensive numerical work exists suggesting a stable 
nontrivial fixed point with $z\approx 5/3$. Thus we predict
from Eq. (\ref{eq:fivetthree}) that the scaling relation 
(\ref{eq:fivefiveb}) holds, with $\zeta=1/z=3/5$ and 
$\chi=4\zeta-2=2/5$ in $d=3$.

Our analysis of the evolution of $ln[c(\x,t)]$ shows in 
effect that the asymptotic behavior of the nonlinear 
equation (\ref{eq:fivetwo}) has the same critical 
exponents as a conventional noisy Burgers equation [23] with 
one less dimension. The same dimensional 
reduction for critical exponents  
$d\rightarrow (d-1)$ was described for the 
original linearized growth model in Sec. IVa. A related 
dimensional reduction has been found by Tang {\it et al.}
\cite{tang} for a model of driven depinning in 
anisotropic media, and by Obuhkov for directed percolation \cite{obuhkov}.

It is interesting  to comment
about the role of $v$ in this perturbative renormalization 
calculation. 
The small dimensionless parameter of the 
series, $g$, is proportional to $1/v$ and 
diverges as $v 
\to 0$.
When $v\to 0$ the theory is  in the strong coupling 
limit where
we expect {\it localized} states in the band tail.
A signal of this phase transition is the (finite) one loop
correction  to $v$ obtained from  Eq. (\ref{eq:fivefourteen})
\be
\delta v \propto -v\int {dq q^{d-2} 
\over (4 D_x D_\perp q^2 + v^2)^{3/2}}.
\eqnum{5.26}
\label{eq:fivetwentysix}
\ee
Although this correction is finite it suggests the existence 
of a critical value of $v$ below which $v_R$ is zero,
consistent with localized states unaffected by convection.

\acknowledgments

It is a pleasure to acknowledge helpful conversations with 
H. Berg, E. Budrene, K. Dahmen, D. Fisher, B.I. Halperin, 
T. Hwa, M. Kardar, A. Lesniewski 
and J. Pelletier. We thank N. Hatano for conversations
and generous help with the figures. Finally, we are indebted
to P. Grassberger for comments and for bringing Refs.
\cite{haarer,movaghar,grassberger}  to our 
attention.  This research was supported
by the National Science Foundation through Grant No. DMR94--17047, 
and by the Harvard Materials Research Science and Engineering 
Laboratory through Grant No. DMR94--00396. One of us (N.S.) 
acknowledges the support of Bar-Ilan University and 
a Rothschild Fellowship.

\appendix 

\section{Population Dynamics and Vortex Configurations}

In this appendix we review the statistical 
mechanics of a superconducting 
vortex line with a columnar pinning 
potential \cite{nel-vin}, and show that 
its partition function evolves with sample 
thickness in the same way as the linearized 
population dynamics problem studied here. The 
different configurations of the vortex line, 
described here simply as an elastic  string, 
are related to possible space-time
trajectories of populations which diffuse, grow and 
drift in an inhomogeneous but time-independent environment. 

Consider a superconductor sample of thickness $L$, pierced by 
``columnar pins,'' which are long aligned 
columns of damaged material,
illustrated schematically in Fig. 11. The 
vortex can usually be described by a single valued 
trajectory $\r(\tau)$, where we assume 
defects are aligned with the $\tau$ direction, 
$\bf{\tau} = \x \times \y$. The free energy of 
this problem may be written as:

\begin{minipage}[t]{3.2in}
\epsfxsize=3.2in
\epsfbox{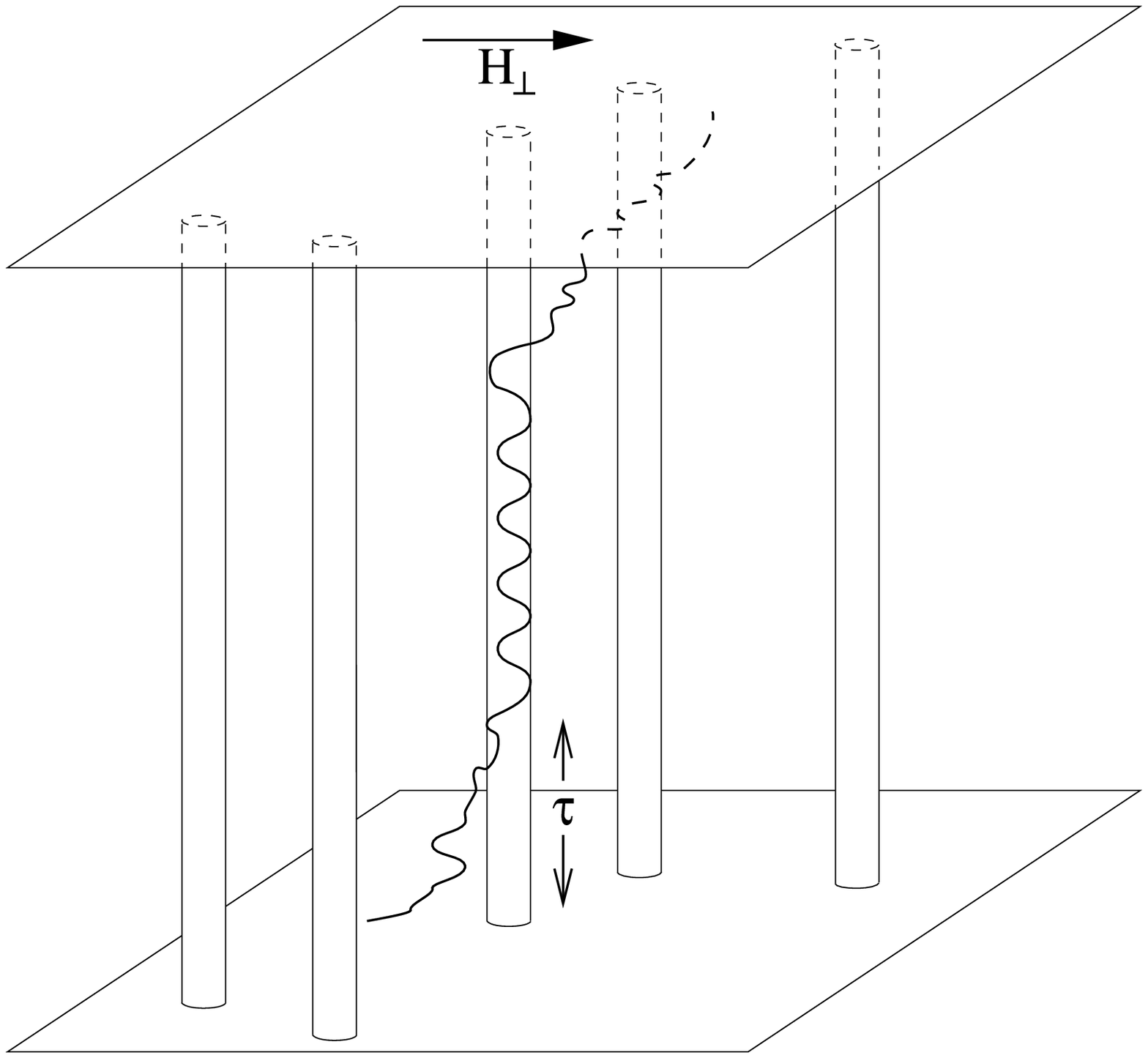}
\begin{small}
FIG.\ 11. 
Vortex line in a superconductor with columnar disorder.
If $H_\perp=0$, the flux line is localized, i.e., trapped 
by one or few pins into some region in the plane 
perpendicular to the correlated disorder. The coordinate
$\tau$ along the columns plays role of time. As the external 
magnetic field is tilted away from the columns, the 
flux line tends to delocalize and tilt in  the 
direction of the external field.
\end{small}
\vspace{0.25in}
\end{minipage}

\noindent
\begin{eqnarray}
F[\r(\tau)] &=& {\epsilon\over 2} 
\int_0^{L} d\tau \left({d\r(\tau) \over 
d \tau}\right)^2
\eqnum{A.1}
\label{eq:aone}  \\
&&+ {1 \over 2}
 \int_0^{L} d\tau \   V[\r(\tau)]  - \H_\perp {\phi_0 \over 4 \pi} 
 \int_0^{L} d\tau
\left( {d\r \over dz} \right)
\nonumber
\end{eqnarray}

where $\epsilon$ is the tilt 
modulus of the flux line and the 
elastic contribution 
${\epsilon \over 2} \left({d\r(\tau) \over \bf{\tau}} \right)^2$ 
is the first nontrivial 
term in the  small tipping angle expansion 
of the line energy of a  nearly
straight vortex line. $V(\r)$ is the random potential which  arises  
from a $\tau$-independent set of 
disorder-induced columnar pinning potentials
(with its average value subtracted off),
and $\H_\perp$ is a perpendicular magnetic  field.

The partition function $\cZ(\x,\tau; \x_0,0)$ 
associated with a vortex which start at position $\x_0$ at the bottom of the 
sample ($\tau \equiv 0)$ and terminates at position 
$\x$ at temperature $T$ somewhere in the interior at 
height $\tau$ is given by a path integral:
\be
\cZ(\x,\tau; \x_0,0) = \int^{\r(\tau) = 
\x}_{\r(0) = \x_0} \cal{D}\r(\tau)
e^{-\cal {F}[\r(\tau)]/T}
\eqnum{A.2}
\label{eq:atwo}
\ee
Standard path integral techniques \cite{nel-dous} 
may be used to show that 
$\cZ(\x,\tau; \x_0,0)$ obeys the Schr\"odinger-like equation,
\be
-T {\partial \cZ \over \partial \tau} =
- {T^2 \over 2 \epsilon} \nabla^2 \cZ
-{T\over\epsilon}{\bf h}_\perp\cdot
{\bf \nabla} Z+ {h_\perp^2\over 2\epsilon}\cZ+
V(\x)Z
\eqnum{A.3}
\label{eq:athree}
\ee
where ${\bf h_\perp} \equiv 
{{\bf H}_\perp  \phi_0 \over 4 \pi}$
is the dimensionless perpendicular field. 
Thus, the growth of the partition function 
$\cZ(\x,\tau)$ of a flexible line with height $\tau$ 
is the same as linearized population growth model  
Eq. (\ref{eq:six}),
with the identifications $\tau \to t$, 
$D \to T/\epsilon$, $\v \to {\bf h}_\perp/
\epsilon$, $U(\x) \to V(\x)/T$ and $a \to h_\perp^2/2 \epsilon T$. 

Because
\be 
\lim_{\tau \to 0} \cZ(\x,\tau; 
\x_0,0) = \delta^d(\x-\x_0)
\eqnum{A.4}
\label{eq:afour}
\ee
the full partition function 
$\cZ(\x,\tau; \x_0,0)$ is in fact the Green's 
function for Eq. (\ref{eq:six}), 
assuming a delta function initial condition 
of population at position  $\x_0$ and time $t = \tau =  0$.

\section{Mean Field Theory of Homogeneous Population Dynamics}

We start with the homogeneous analogue of Eq. (\ref{eq:two}),
\begin{eqnarray}
{\partial \over  \partial t} c(\x,t) + \v \cdot {\bf \nabla} c(\x,t)
&=& 
\eqnum{B.1}
\label{eq:bone}  \\ 
 &D& \nabla^2 c(\x,t) + a c(\x,t) - b c^2(\x,t)  \nonumber 
\end{eqnarray}
Upon decomposing $c(\x,t)$ into Fourier modes
\be
c(\x,t) = {1 \over \Omega} \sum_\k c_\k(t) e^{i\k\cdot\x}
\eqnum{B.2}
\label{eq:btwo}
\ee
where $\Omega$ is the volume of a 
$d$-dimensional box with periodic boundary 
conditions, we have
\be
{dc_\k(t) \over dt} = \Gamma_\k c_\k - 
{b \over \Omega} \sum_{{\bf q,q'}}
c_\q(t) c_{\q'}(t) \delta_{\k,\q+\q'}
\eqnum{B.3}
\label{eq:bthree}
\ee
with the complex growth spectrum
\be
\Gamma_\k = a + i \v \cdot \k - D k^2.
\eqnum{B.4}
\label{eq:bfour}
\ee
In the spirit of Bogoliubov approximation for the 
collective excitations in 
superfluids \cite{tang}, we separate out the $\k=0$ mode
and write
\be
{dc_0(t) \over dt} = a c_0 - {b \over \Omega}  
c_0^2 - \left[ {b \over \Omega}
\sum_{\q \neq 0} 
c_\q(t) c_{\k-\q}(t) \right]   
\eqnum{B.5}
\label{eq:bfive}
\ee
\begin{eqnarray}
{dc_\k(t) \over dt} = \Gamma_\k c_\k &-& {2b 
\over \Omega} c_0(t) c_\k(t) \nonumber \\ &-&
\left[ {b \over \Omega} \sum_{\q \neq 0, \q'  \neq \k   }
c_\q(t) c_{\k-\q'}(t) \right]
\eqnum{B.6}
\label{eq:bsix}
\end{eqnarray}

The mean field approximation consists of neglecting 
terms in the modes with 
$\k \neq 0$, shown in brackets in Eqs. 
(\ref{eq:bfive}) and (\ref{eq:bsix}).
The approximate differential equations which remain have solutions
\be 
c_0(t) = {\Omega\bar{c}_0 e^{at} \over 
[1+ {b \bar{c}_0  \over a} (e^{at}-1)]}
\eqnum{B.7}
\label{eq:bseven}
\ee

\be
c_\k(t) = { c_\k(0) e^{\Gamma_k t}  \over  
 [1+ {b \bar{c}_0  \over a} (e^{at}-1)]^2 }
\eqnum{B.8}
\label{eq:beight}
\ee
where $\bar{c}_0=\Omega c_0(t=0)  $. 
The time evolution
of $c(\x,t)$ is now completely determined 
by inserting these results into 
Eq. (\ref{eq:btwo}). Note that there are 
many growing modes with $\k \neq 0$ 
at short times when $a>0$. However, when $t 
\gg \ln(b \bar{c}_0/a)$ the 
denominators of Eqs. (\ref{eq:bseven}) and (\ref{eq:beight}) cause 
$c_0(t)$ to approach $c_0
= a/b$, and all modes with $\k \neq 0$ to 
decay away with the spectrum 
(\ref{eq:sixteen}).

\section {DENSITY OF STATES NEAR THE BAND EDGE}

In this appendix we calculate the density of states (DOS) in the 
tail of the band of growth rates, i.e., near the ground state. 
We
set the growth bias $a\equiv 0$, since it does not affect the 
statistics of the DOS. The discussion below 
assumes localized eigenmodes in  
the ``relevant'' or unstable part of the  spectrum, where 
the  convection term  $\g\propto \v$  just produces a 
trivial shift in  the 
eigenvalues, and hence we neglect $\g$ as well.

Consider, then,   
a $d$-dimensional hypercubic lattice with edge length $\ell_0$,  
where the potential energy 
at each site is in the range $U(\x)\in [-\Delta,\Delta]$.
Unbounded  probability measures may give different results.

The continuum Liouville operator approximated by the lattice
model may be written as:
\be
\cL =  D\nabla^2 +U(\x). 
\eqnum{C.1}
\label{eq:cone}
\ee
It is easy to see that the DOS function 
$\rho(E)$  in this model is  bounded
from above  by  $\Delta$, such that $\rho(E) \to 0$ 
as $E \to \Delta$.  
The tail of the DOS  is  
determined by the range of energies in which 
the DOS is 
determined by rare events, characterized by   
large spatial regions with 
low potential energy \cite{friedberg}.  

Let us estimate these fluctuations in 
the following way:  the probability to 
find a hypersphere of radius $R$ which  contains 
only blocks of potential energy $U$ larger  
than $\Delta-V_0$ is approximately  
\begin{eqnarray}
P(R, U > \Delta - V_0) & \approx & 
\left({V_0\over 2\Delta}\right)^{(R/\ell_0)^d}
\nonumber \\ 
& \approx & \exp\left[\left({R\over \ell_0}
\right)^d\ln \left({V_o\over 2\Delta}
\right)\right].
\eqnum{C.2}
\label{eq:ctwo}
\end{eqnarray}
The energy of a state confined by this rare fluctuation is 
given approximately by
\be 
E \approx \Delta -V_0-D/R^2,
\eqnum{C.3}
\label{eq:cthree}
\ee
so that the probability to get an energy between $E$ and
$E+dE$
using a sphere of radius $R$
is $p(R,E)\sim{\partial P\over\partial V_0}|_{V_0=E}$, i.e., 
\be
p(R, E) \sim \exp\left[
\left( {R\over l_0}\right)^d
ln\left({1\over 2}-{E\over 2\Delta}-{D\over 2R^2\Delta}
\right)\right].
\eqnum{C.4}
\label{eq:cfour}
\ee
This expression is well defined for 
$\sqrt {{D\over \Delta-E}} \leq R < \infty  $.
Optimizing $p(R, E)$ with respect to 
$R$ gives, up to logarithmic 
corrections, a  maximum at the lower limit  
$R^* \approx  \sqrt{ D/(\Delta-E)}$, so that as $E \to \Delta$
from below, $p(E)\equiv p(R^*,E)$ vanishes according to  
\be
p_(E) \sim \exp[-(D/\ell_0^2[\Delta-E])^{d/2}].
\eqnum{C.5}
\label{eq:cfive}
\ee
Now  the DOS is proportional to $p(E)$, i.e., at the tail
of the distribution we have  
$g(E) \sim g_0 p(E)$, where $g_0$ is some normalization
of order  the DOS in the middle of the band.

Let us consider now the  tight-binding analog of the above model. 
The on site potential is  $U(\x)$, taken 
from a  square distribution in the range $[-\Delta,\Delta]$.
The Liouvillian is, from Eq. (\ref{eq:twenty}),
\begin{eqnarray}
\tilde \cL &=& {w \over 2} \sum_\x \sum_{\nu = 1}^d  
[|\x\rangle\langle\x+\e_\nu|\;+\; |\x + \e_\nu\rangle
\langle\x|]\nonumber \\ &&+ 
\sum_x U(\x)|\x\rangle\langle\x|\;.
\eqnum{C.6}
\label{eq:csix}
\end{eqnarray}
The eigenenergies of this Hamiltonian are  
bounded, $-\Delta-w < \epsilon_n < \Delta +w$ with $w\sim D/\ell_0^2$.
The states in the tails correspond to  rare spatial fluctuations 
of $U(\x)$. The probability to find 
such fluctuations (e.g., a region
of radius $R$ in which the on site 
potential is within a specified energy interval $V_0$
of the maximum value $\Delta$) is 
the same as in the previous model. The  
energy spectrum of such fluctuation
is given approximately by
\be
\epsilon = \Delta -V_0 + w \sum_{\nu = 1}^d \cos(k_\nu\ell_0),
\eqnum{C.7}
\label{eq:cseven}
\ee
where $k\approx k_{\min} \sim 1/R$. Thus, states in this tail
obey the relation
\be
\epsilon \sim \Delta+w-V_0-w/R^2
\eqnum{C.8}
\label{eq:ceight}
\ee
so that a result of the form (C.5) is  
applicable 
here also, with the energy  measured from the edge of the band
defined by the lattice model \cite{evmodel}.

We can use the density of states result (C.5) and the eigenvalue spectra 
displayed in Figs. 2 to understand the results of Refs. \cite{movaghar} and 
\cite{grassberger} for particles diffusing and drifting with random traps 
in one dimension. We start with the expansion of $c(\x,t)$ in the complete 
set of eigenfunctions of Eq. (1.2) with $b=0$, as in Eq. (2.1). For ${\bf
v = g} =0$, all eigenvalues $\Gamma_n$ are real and negative for this problem. 
As discussed in \cite{grassberger}, the  eigenvalues close to zero 
arise from the rare regions discussed above for the density of states. 
The spectrum for the lattice model will look like Fig. (2a), with 
however the top of the band just touching the origin. Upon assuming 
a uniform initial condition $c(\x,0)$, we take 
all $c_n \sim  const$ and integrate over space to obtain
\be
\eqnum{C.9}
N_{tot} = \int d^d\x c(\x,t) \propto \int_{-\infty}^0 g(\Gamma) e^{\Gamma t}
 d\Gamma
\ee
Where $ g(\Gamma)$ is the density of states. At $t \to \infty$, 
the behavior of the total number of surviving particles is 
dominated by energies near the top of the band, where 
\be  
\eqnum{C.10}
g(\Gamma) \propto \exp[-|\Gamma_0/\Gamma|^{d/2}].
\ee
A saddle point evaluation of Eq. (C.9) in $d$ dimensions
then leads to
\be
\eqnum{C.11}
N_{tot}(t) \propto \exp[-(t/t_0)^{d \over d+2}],
\ee
in agreement with Refs. \cite{movaghar} and \cite{grassberger}.

For ${ v = g < g_1}$, the effect of nonzero 
convection is simply a rigid downward shift of the spectrum, 
by an amount of $v^2/4D$ \cite{grassberger}. When ${ g > g_1}$ in 
one dimension, the bubble of delocalized states shown in fig. (2b) will appear 
in the center of the band of negative eigenvalues. However, 
the behavior of the density of states at the top 
of band is unchanged, and we find
\be
\eqnum{C.12}
 N_{tot}(t)  \propto \exp[- {v^2 t \over 4 D}] \exp[-(t/t_0)^{1/3}]
\ee
in agreement with Ref. \cite{movaghar}. Delocalization affects the long time 
decay only when ${ g>g_2}$, when {\it all} states are delocalized, as in Fig. 
2c. The density of states in one dimension is then
\be
\eqnum{C.13}
g(\Delta \Gamma) \propto 1/(\Delta \Gamma)^{1/2}
\ee
which leads an additional exponential contribution to the decay of 
Eq. (C.9) (up to logarithmic corrections), in qualitative 
agreement with the transition as a 
function of the bias found by Movaghar et. al. \cite{movaghar}.
It would be interesting to use the spectra 
displayed in Figs. 2 and 3 
to  extract the short and intermediate 
time behavior of $N_{tot}$, as well as the effect of drift in higher 
dimensions.

 Note that the delocalization transition which describes  particles 
diffusing and convecting in the presence of traps occurs 
at the {\it top} of the band. For population biology problems, one must 
consider {\it positive} growth eigenvalues and the phenomena of interest 
typically occur for ${ g_1 < g < g_2}$. The localization
transition of interest to us in this paper occurs when the {\it 
mobility edge} crosses the origin, as in Fig. 5.

\end{multicols}{2} 
\end{document}